%% file: arxiv-main.tex
\documentclass[11pt]{article}

\usepackage[a4paper,margin=2.3cm]{geometry}
\usepackage[T1]{fontenc}
\usepackage[utf8]{inputenc}
\usepackage{libertinus}

\usepackage[nopatch=eqnum]{microtype}

\usepackage{amsmath,amssymb,amsfonts,amsthm,mathrsfs}

\usepackage{graphicx}
\usepackage{subcaption}
\usepackage{multirow}
\usepackage{booktabs}
\usepackage{tabularx}
\usepackage{wrapfig}
\usepackage{siunitx}
\usepackage{textcomp}
\usepackage[svgnames]{xcolor}
\usepackage{caption}
\usepackage{epstopdf}

\usepackage{algorithm}
\usepackage{algorithmicx}
\usepackage{algpseudocode}
\usepackage{listings}

\usepackage{authblk}

\usepackage{titlesec}

\usepackage{titlesec}

\titleformat{\section}
  {\large\bfseries}
  {\thesection.}
  {0.6em}
  {}

\titleformat{\subsection}
  {\normalsize\bfseries}
  {\thesubsection.}
  {0.5em}
  {}

\titleformat{\subsubsection}
  {\normalsize\bfseries\itshape}
  {\thesubsubsection.}
  {0.5em}
  {}

\titleformat{\paragraph}[runin]
  {\normalsize\bfseries}
  {}
  {0pt}
  {}

\titlespacing*{\section}
  {0pt}
  {1.8ex plus 0.4ex minus 0.2ex}
  {0.7ex plus 0.2ex}

\titlespacing*{\subsection}
  {0pt}
  {1.3ex plus 0.3ex minus 0.2ex}
  {0.5ex plus 0.1ex}

\titlespacing*{\subsubsection}
  {0pt}
  {1.1ex plus 0.3ex minus 0.1ex}
  {0.7ex plus 0.1ex}

\titlespacing*{\paragraph}
  {0pt}
  {1.0ex plus 0.2ex minus 0.1ex}
  {0.7em}

\theoremstyle{plain}

\theoremstyle{definition}

\usepackage[numbers,sort&compress]{natbib}
\usepackage[colorlinks=false,pdfborder={0 0 0}]{hyperref}

\input{math_commands}

\title{A differentiable machine learning small-angle X-ray scattering analysis framework for structure elucidation of lipid nanoparticles}
\author[1,]{Maria Bånkestad\thanks{Corresponding author: \texttt{maria.bankestad@ri.se}}}
\author[2,]{Sandra Barman}
\author[2,3,4]{Magnus Röding}
\author[2,3]{Erik Kaunisto}
\author[5]{Viktoriia Meklesh}
\author[5]{Audrey Gallud}
\author[5]{Marco Mendez}
\author[5]{Marianna Yanez Arteta}
\author[6]{Stefan Norberg}
\author[7]{Ann Terry}
\author[1]{Smita Chakraborty}
\author[8]{Shun Yu}
\author[8]{Jerk Rönnols}
\author[1]{Sepideh Pashami}

\affil[1]{RISE Research Institutes of Sweden, Division Digital Systems, Computer Science, Kista, Sweden}
\affil[2]{RISE Research Institutes of Sweden, Division Bioeconomy, Food Research and Innovation, Göteborg, Sweden}
\affil[3]{Sustainable Innovation \& Transformational Excellence, Pharmaceutical Technology \& Development, Operations, AstraZeneca Gothenburg, Mölndal, Sweden}
\affil[4]{Department of Mathematical Sciences, Chalmers University of Technology and University of Gothenburg, Göteborg, Sweden}
\affil[5]{Advanced Drug Delivery, Pharmaceutical Sciences, R\&D, AstraZeneca, Mölndal, Sweden}
\affil[6]{Global Product Development, Pharmaceutical Technology \& Development, Operations, AstraZeneca Gothenburg, Mölndal, Sweden}
\affil[7]{MAX IV Laboratory, Lund University, Lund, Sweden}
\affil[8]{RISE Research Institutes of Sweden, Division Bioeconomy, Sustainable Materials and Packaging, Stockholm, Sweden}

\date{}

\begin{document}

\maketitle

\begin{abstract}
\noindent Lipid nanoparticles (LNPs) are efficient delivery systems for negatively charged nucleic acids. Their multi-component architecture yields a core--shell structure. Small-angle X-ray scattering (SAXS) is an important characterization technique for LNPs, but recovering internal structure and size distribution from SAXS is an inverse problem with non-unique solutions. Realistic models are often too expensive for systematic exploration. We introduce a machine-learning-accelerated, differentiable framework for SAXS analysis of heterogeneous, polydisperse LNPs. The forward model combines a core--shell particle with a Gaussian-random-field interior, a neural surrogate for the monodisperse SAXS map, and a differentiable layer integrating over particle-size distributions. The surrogate reduces prediction cost by four orders of magnitude, while differentiability enables large-scale multi-start fitting and ensemble identifiability analysis. Applied to synthetic and experimental MC3 LNP data, the framework shows that near-identical SAXS fits can arise from distinct parameter modes, with the experimental fits dominated by a trade-off between size-distribution and interior-structure parameters.

\end{abstract}


\section{Introduction}\label{sec:intro}


Lipid nanoparticles (LNPs) have enabled major advances in nucleic acid delivery, including siRNA therapeutics (Onpattro\textsuperscript{\textregistered}), mRNA vaccines (Comirnaty\textsuperscript{\textregistered}, Spikevax\textsuperscript{\textregistered}), and therapeutic gene-editing programs now in clinical trials
\cite{Padilla2025elucidating, aliakbarinodehi2024time, hou2021lipid}.
Conventionally, LNPs are prepared using microfluidic mixing, which yields particles with narrow size distributions and high encapsulation efficiencies. The resulting particles are generally spherical and are often described as exhibiting a core--shell structure \cite{cullis2017lipid, cardenas2023review}. However, the processes governing LNP formation and RNA packing remain incompletely understood. In practice, the interior lipid organization and the spatial distribution of the cargo can vary, the shell may differ in thickness and composition \cite{caselli2024small}, and particle populations are typically polydisperse in size and shape \cite{hou2021lipid, kulkarni2018lipid, dehghani2023ionizable}. These structural features matter because interior structure, size distribution, and shape collectively influence particle stability, biodistribution, and therapeutic performance.

Characterizing LNP structure requires complementary measurements \cite{hallan2021challenges}.
Dynamic light scattering (DLS) estimates particle size distributions from diffusion but reports a hydrodynamic radius that can be biased by polydispersity, non-spherical geometry, or non-ideal refractive index \cite{stetefeld2016dynamic}.
Cryogenic transmission electron microscopy (cryo-TEM) provides direct images of individual particles and resolves structural variability at the single-particle level \cite{cardenas2023review}, but requires vitrification and may not be representative of the native dispersion.
Small-angle X-ray scattering (SAXS) is a powerful method: it can be performed in solution without extensive sample preparation, and it provides population-averaged reciprocal-space information across length scales from a few ångström to hundreds of nanometers \cite{li2016small,hammel2023correlating}.

For a solution of randomly oriented particles, SAXS records an isotropic two-dimensional scattering pattern, which is radially averaged to a one-dimensional intensity profile $I(q)$, where $q$ is the magnitude of the scattering vector. This profile encodes contributions from particle geometry, size distribution, and interior electron-density variations, all mixed together via a Fourier transform.
Because of incomplete and averaged information, many distinct three-dimensional electron-density configurations can produce nearly identical one-dimensional SAXS profiles \cite{pedersen2014quantification,taylor2003phase, do2020small, archibald2020classifying}.
This non-uniqueness is inherent to SAXS measurements, not an artifact of noise or a limited model specification.
Meaningful interpretation of SAXS data, therefore, requires a forward model informed by prior knowledge and complementary measurements, with inferred parameters understood as conditional on that model \cite{narayanan2017recent, glatter1982small}.

For heterogeneous, polydisperse systems such as LNPs, realistic models are computationally expensive. Analytical models are fast but too idealized for meaningful interpretation across all length scales. Numerical models that resolve the three-dimensional electron density and average over stochastic realizations can capture interior variability \cite{pedersen2012structure, schmidt2007simulation, roding2022, kronenberger2024random}, but each forward evaluation is costly, and polydispersity compounds this cost by requiring evaluation across a distribution of radii at every fitting step.
This limits the practical usefulness of realistic SAXS models \cite{penttila2021combining, philipp2025combining, lytje2025small}. The deeper consequence is not merely slow fitting: without a fast, differentiable model, it is not feasible to systematically investigate the parameter space for a model that incorporates both interior structure and size distribution.
Machine learning has been used to accelerate SAXS analysis for liquid crystalline phases \cite{aty2022machine}, nanoparticle shape and size classification \cite{monge2024automated, roberts2025automated, adhikari2025quantifying},
signal denoising \cite{zhou2023machine}, and surrogate-accelerated parameter recovery for colloidal systems \cite{wong2024predicting}, but to our knowledge, there is no approach that combines complex morphological models for LNPs with machine learning acceleration.

\begin{figure}[t!]
    \centering
    \includegraphics[width=0.99\linewidth]{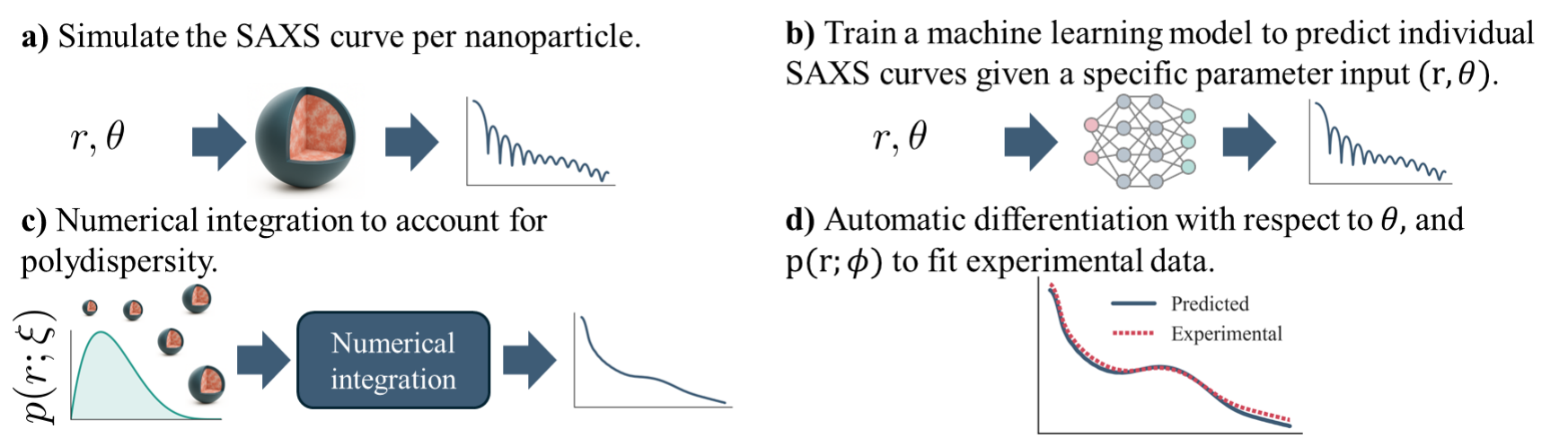}
    \caption{
    \textbf{Overview of the SAXS inference framework.}
    \textbf{(a)} A stochastic core--shell density model with a GRF core defines a monodisperse forward simulator.
    \textbf{(b)} A neural surrogate approximates the monodisperse forward map on a fixed \(q\) grid.
    \textbf{(c)} Polydispersity is incorporated via a differentiable radius distribution and fixed-quadrature evaluation of the ensemble-averaged intensity.
    \textbf{(d)} Experimental profiles are fitted by gradient-based multi-start optimization; retained near-optimal solutions are used for trade-off and sensitivity analyses.
    }
    \label{fig:model_overview}
\end{figure}

In this work, we develop a machine learning--accelerated, differentiable inference framework for SAXS analysis of heterogeneous, polydisperse lipid nanoparticles (Fig.~\ref{fig:model_overview}) and use it to characterize what a one-dimensional SAXS profile can and cannot jointly determine about LNP structure and size distribution.
The starting point is a monodisperse forward model with a homogeneous shell and a heterogeneous core represented as a Gaussian random field (GRF).
Given structural parameters, the forward map samples a realization of the interior field, embeds it within the core--shell geometry, and computes the resulting orientationally averaged SAXS intensity.
The GRF serves as a statistical model of interior heterogeneity rather than a mechanistic account of lipid organization \cite{rasmussen2006gaussian, adler2007grf, berk1991scattering, gommes2013three, gommes2018stochastic, gommes2020stochastic}, and its spectral parameters are interpretable as a distribution of characteristic length scales. 
We train a neural surrogate of this monodisperse forward map, which reduces the cost of a SAXS prediction by roughly four orders of magnitude.
A differentiable polydispersity layer then integrates the surrogate over a truncated log-normal radius distribution, making the full polydisperse forward model both fast and differentiable with respect to structural and size-distribution parameters.
This combination makes large-scale multi-start gradient-based fitting practical, so that hundreds of independent optimizations run in under a minute on a single GPU, a task that would otherwise take several days.
We apply the framework to synthetic polydisperse benchmarks with known ground truth and to experimental SAXS data from Dlin-MC3-DMA (MC3) LNPs, a well-studied system for siRNA therapeutics \cite{hou2021lipid,philipp2023ph}.
The resulting near-optimal ensembles reproduce the scattering profile closely across the measured $q$-range in both settings. Obtaining this quality of fit for a polydisperse, heterogeneous LNP is itself a non-trivial result, and it is what allows structure and size distribution to be analyzed within a single model. Close agreement to the profile, however, does not imply unique recovery of the underlying parameters.
For the experimental data, the ensemble separates into physically distinct modes, with a dominant structured trade-off between size-distribution parameters and interior-structure parameters, leaving the curve shape essentially unchanged.
Gradient-based sensitivity analysis characterizes this trade-off directly from the surrogate's Jacobians, and a comparison with analytical homogeneous-core models identifies the regime in which the heterogeneous model is needed.

We present synthetic and experimental results in Section~\ref{sec:results}, followed by discussion in Section~\ref{sec:discussion} and methods in Section~\ref{sec:methods}.

\section{Results}
\label{sec:results}

We first benchmark the inference pipeline on synthetic data with known parameters, then apply it to experimental MC3 LNP data, where we analyze fit quality, parameter distributions, and sensitivity. We then compare with analytical baseline models and report surrogate accuracy and computational cost.
  
\subsection{Synthetic benchmarks: recovery and identifiability}
\label{sec:synth_fit}

\begin{figure}[b!]
    \centering
    \includegraphics[width=\linewidth]{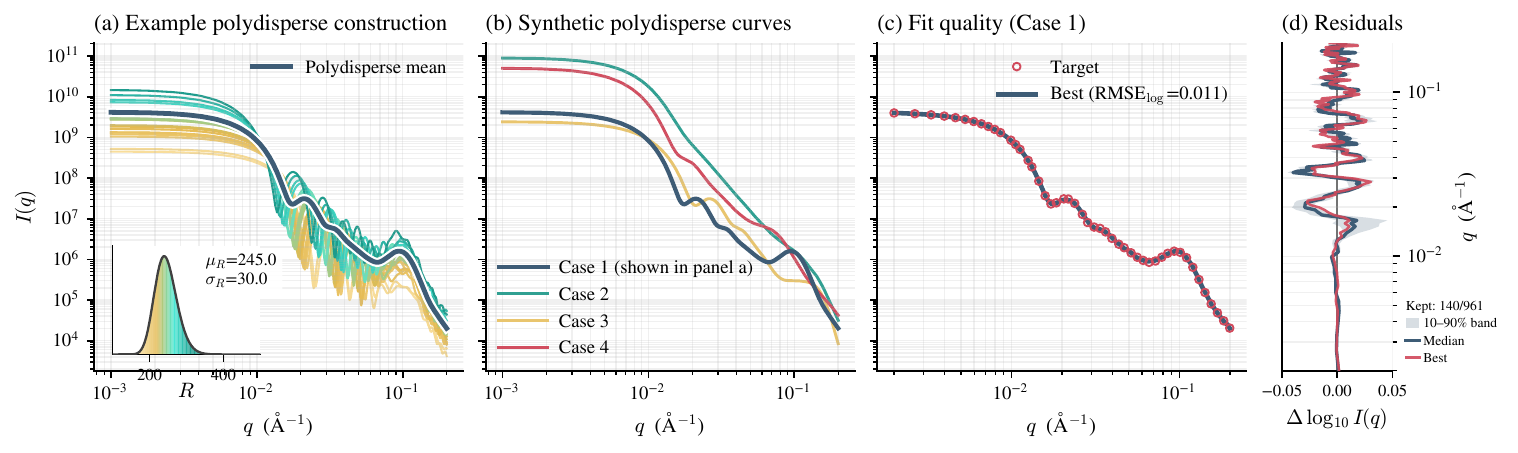}
    \caption{
    Synthetic polydisperse SAXS targets and representative fit quality.
    \textbf{(a)} Construction of a polydisperse curve from radius-resolved contributions (colored by radius) and the resulting Monte Carlo mean (thick line); inset: truncated log-normal radius distribution.
    \textbf{(b)} Four synthetic polydisperse SAXS targets generated in the same way with different radius and/or structural parameter distributions; the target shown in (a) is highlighted.
    \textbf{(c)} Fit quality for Case~1: target polydisperse SAXS curve (markers) and the best fit among the retained set (line). All retained solutions are plotted but produce curves indistinguishable from the best fit at this scale; their dispersion is shown in panel~(d).
    \textbf{(d)} Log-intensity residuals relative to the target, $\Delta\,\log_{10} I(q)=\log_{10} I_{\mathrm{pred}}(q)-\log_{10} I_{\mathrm{true}}(q)$: the shaded band shows the 10--90th percentile across retained solutions, the dark line is the median residual, and the red line is the best-solution residual. All errors are computed in $\log_{10} I(q)$ space.
    }
    \label{fig:fig_synth_overview_and_fit}
\end{figure}

To assess what the SAXS inverse problem can and cannot resolve under controlled conditions, we benchmark the inference pipeline on four synthetic polydisperse targets with known ground-truth parameters (Methods, section~\ref{sec:synth_targets}).
Each target is constructed by Monte Carlo averaging over $N=5000$ independent particle simulations, making residual stochastic variability negligible.
The four targets span a range of curve shapes across the $q$-range (Fig.~\ref{fig:fig_synth_overview_and_fit}b), reflecting different combinations of size-distribution and interior-structure parameters (Supplementary Table~S2).

We fit each target using multi-start gradient-based optimization and retain an ensemble of near-optimal solutions (Methods, section~\ref{sec:eval_metrics}).
A representative case (Case~1) is shown in Fig.~\ref{fig:fig_synth_overview_and_fit}c--d: many distinct parameter settings produce visually indistinguishable curves with small residuals across the full $q$-range.

\begin{figure}[t!]
    \centering
    \includegraphics[width=0.95\linewidth]{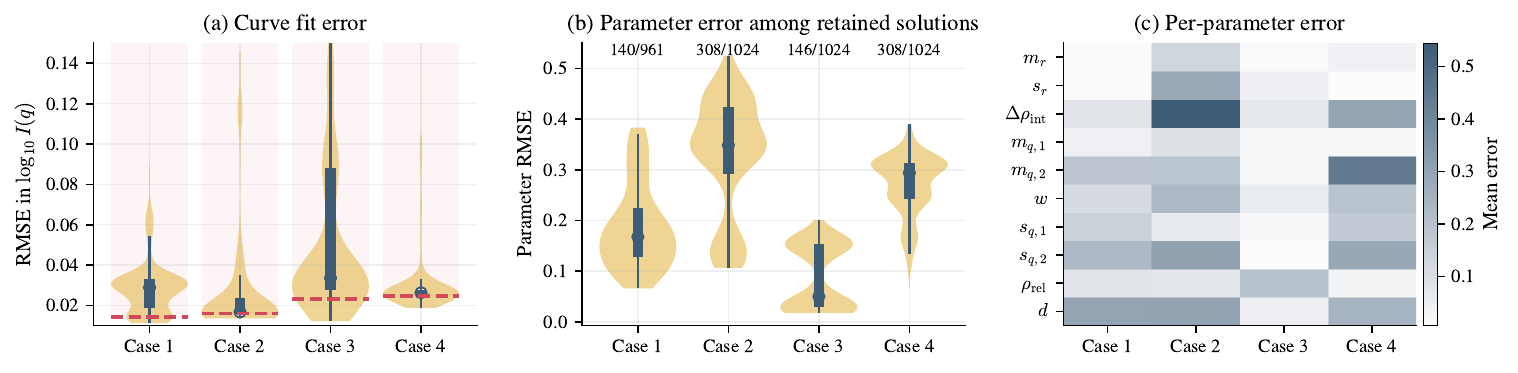}
    \caption{
    Synthetic inverse-problem outcomes across four polydisperse targets.
    (a) Distribution of final curve RMSE in $\log_{10} I(q)$ over optimization runs for each synthetic case; dashed lines indicate the per-case retention thresholds.
    (b) Distribution of parameter RMSE in min--max normalized space for the retained solutions, showing that excellent curve fits can correspond to diverse parameter sets.
    (c) Mean absolute error per parameter (min--max space), aggregated over retained solutions, highlighting which parameters are well constrained and which remain degenerate.
    }
    \label{fig:fig_synth_error_summary}
\end{figure}

Across all four targets, the retained ensembles achieve low curve error in $\log_{10} I(q)$ space (Fig.~\ref{fig:fig_synth_error_summary}a), confirming that the surrogate-based forward model can accurately reproduce the target scattering.
Parameter recovery is a different matter.
Even within the near-optimal sets, normalized parameter errors are often large and structured (Fig.~\ref{fig:fig_synth_error_summary}b--c): the polydisperse SAXS inverse problem admits multiple solutions, and multiple parameter settings can produce effectively the same scattering profile even without measurement noise.

Good fit quality in curve space is therefore not a reliable indicator of parameter recovery.
Case~3 achieves a comparatively low parameter error despite a slightly higher curve RMSE. The other cases show that low curve error does not guarantee low parameter error: their fits are near-indistinguishable, yet the underlying parameter values differ substantially.
The dominant ambiguity is a compensatory trade-off between size-distribution parameters $(m_r, s_r)$ and GRF spectrum parameters (e.g., $m_{q,1}$, $w$, $s_{q,1}$): adjusting the particle-size statistics can be partially offset by changes in the interior spectrum while preserving the broad features of the curve. Parameter definitions and bounds are listed in Table~\ref{tab:param_bounds} in the methods section.

The per-parameter error profiles (Fig.~\ref{fig:fig_synth_error_summary}b,c) give a more specific view. The overall parameter RMSE (panel b) ranges from about $0.04$ for Case~3, where recovery is tight across all parameters, up to roughly $0.16$--$0.42$ for Cases~1, 2, and~4. Panel~(c) resolves this into individual parameters and shows that the high-error cells are case-specific: Case~2 peaks at $\Delta\rho_{\mathrm{int}}$, and Case~4 at $m_{q,2}$.
No single parameter is loose in every case, and apart from Case~3 no single parameter is tight in every case either. We now ask whether the same degeneracy structure appears in experimental data, where the ground truth is unknown.

\subsection{Experimental SAXS analysis of MC3 LNPs}
\label{sec:exp_fit}

\begin{figure}[t!]
    \centering
    \includegraphics[width=\linewidth]{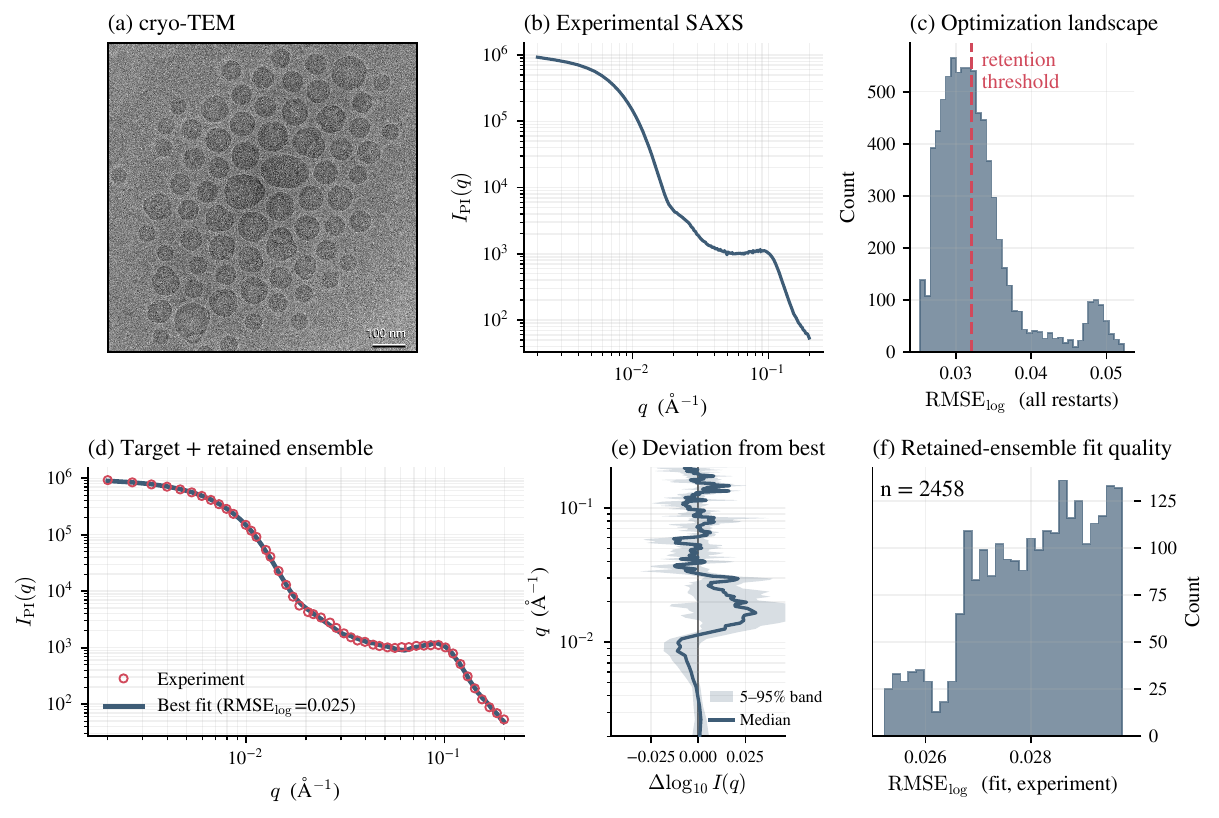}
    \caption{%
    \textbf{Experimental MC3 LNPs: characterization, optimization, and ensemble fit quality.}
    \textbf{(a)} Cryo-TEM micrograph of MC3 LNPs in vitrified ice (200\,kV, JEOL JEM-2100Plus); manual annotation of 400 particles yields a mean radius of $346 \pm 62$~\AA.
    \textbf{(b)} Experimental SAXS profile on a log--log scale, covering the analysis $q$ range ($0.001$--$0.200$~\AA$^{-1}$).
    \textbf{(c)} Distribution of $\mathrm{RMSE}_{\log}$ across 8000 optimization restarts; the dashed line marks the retention threshold (the histogram clips the top 4\% of restarts for visibility).
    \textbf{(d)} Experimental target (markers), best fit (line), and retained ensemble plotted on log axes; the retained curves collapse onto the best fit within line width.
    \textbf{(e)} Per-$q$ deviation $\log_{10} I_{\mathrm{fit}}(q) - \log_{10} I_{\mathrm{best}}(q)$ across the retained ensemble: median (line) and $5$--$95$\,\% band (shaded).
    \textbf{(f)} Distribution of $\mathrm{RMSE}_{\log}$ between retained fits and the experimental SAXS curve.
    }
    \label{fig:exp_overview}
\end{figure}

We fit an experimental SAXS profile from MC3 LNPs (see Methods, section~\ref{sec:experimental} for sample preparation and measurement details).
Figure~\ref{fig:exp_overview}a--b shows the sample imaged by cryo-TEM alongside the experimental SAXS curve.
Manual annotation of 400 particles in the cryo-TEM micrograph gives a radius distribution of $346 \pm 62$~\AA, and DLS yields a (hydrodynamic) radius distribution of $323 \pm 79$~\AA{} (note that for DLS we also report the number-weighted distribution).
Both provide independent size estimates but measure different effective radii: DLS analysis yields a hydrodynamic radius that generally overestimates the true radius, while cryo-TEM provides information on LNPs in a vitrified state that may not be representative of LNPs in solution.
Neither value is directly comparable to the radius distribution inferred from SAXS, though both provide independent order-of-magnitude checks on particle size.

We perform 8000 independent gradient-based optimizations from Sobol-sampled initializations and retain a near-optimal ensemble of low-error solutions (Methods, section~\ref{sec:eval_metrics}).
The distribution of final $\mathrm{RMSE}_{\log}$ across restarts (Fig.~\ref{fig:exp_overview}c) shows a wide spread, with many restarts converging to substantially higher loss values, reflecting a non-convex objective landscape with multiple local minima and saddle regions.
The retained low-error subset forms the basis for all subsequent analyses.

The retained ensemble reproduces the experimental SAXS profile across the full $q$-range (Fig.~\ref{fig:exp_overview}d--f).
The retained curves collapse onto the best fit within line width (Fig.~\ref{fig:exp_overview}d), and the median deviation from the best fit, together with its $5$--$95$\,\% band, stays small across $q$ (Fig.~\ref{fig:exp_overview}e).
The distribution of $\mathrm{RMSE}_{\log}$ against the experimental curve over the retained set (Fig.~\ref{fig:exp_overview}f) confirms that all retained solutions, not only the single best restart, achieve comparably good fits to the data.
The curve shape is tightly constrained by the data; the underlying parameter values, as the following sections show, are not.

\subsubsection{Parameter distributions and trade-offs}
\label{sec:param_tradeoffs}

\begin{figure}[t!]
    \centering
    \begin{subfigure}[t]{0.99\linewidth}
        \centering
        \includegraphics[width=\linewidth]{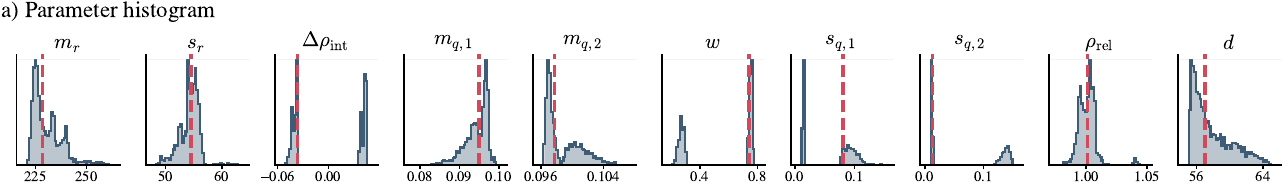}
    \end{subfigure}\vfill
    \vspace{0.2cm}
    \begin{subfigure}[t]{0.23\linewidth}
        \centering
        \includegraphics[width=\linewidth]{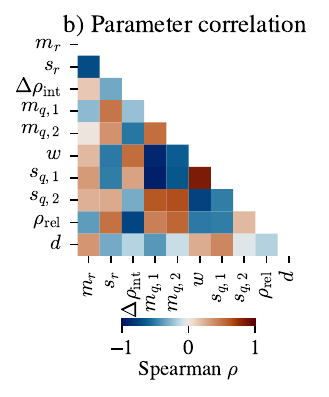}
    \end{subfigure}\hfill
    \begin{subfigure}[t]{0.7699\linewidth}
        \centering
        \includegraphics[width=\linewidth]{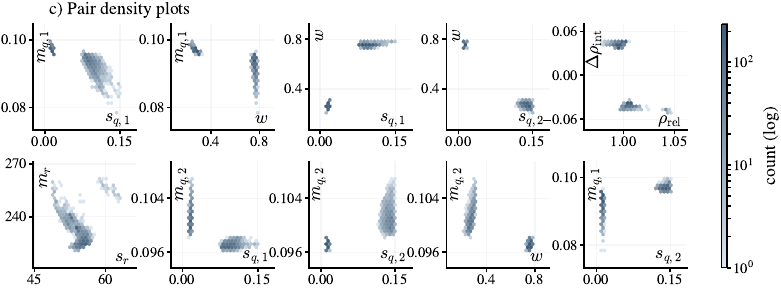}
    \end{subfigure}
    \caption{
    Correlation structure of the inferred parameters for the retained near-optimal solutions.
    (a) Marginal parameter histograms across optimization restarts.
    (b) Spearman rank-correlation matrix, highlighting monotonic dependencies and parameter trade-offs.
    (c) Hexbin density plots for the ten parameter pairs with the largest absolute Spearman correlation, revealing structured manifolds in parameter space.
}
    \label{fig:correlation}
\end{figure}

Although the retained curves are nearly indistinguishable, the inferred parameters vary in a structured way (Fig.~\ref{fig:correlation}).
Several marginal histograms are broad and clearly multimodal (Fig.~\ref{fig:correlation}a), showing that the experimental SAXS profile does not uniquely determine all structural degrees of freedom, even within the near-optimal set.\looseness-1

The rank-correlation matrix (Fig.~\ref{fig:correlation}b) reveals strong monotonic dependencies consistent with compensatory trade-offs in the forward model.
The GRF spectrum parameters $s_{q,1}$ and $w$ are strongly positively correlated, while $w$ is strongly anticorrelated with $m_{q,1}$.
Pronounced coupling also appears between the size-distribution parameters $(m_r, s_r)$ and the interior spectrum parameters (e.g., $m_{q,1}$), indicating that changes in particle-size statistics can be partially offset by changes in the interior spectrum while producing a similar intensity profile.

The pairwise projections (Fig.~\ref{fig:correlation}c) show these trade-offs as elongated, curved manifolds.
Several of the strongest parameter pairs exhibit two distinct density regions rather than a single connected cloud, consistent with multiple parameter modes that yield comparably good fits.
The data constrain certain parameter combinations more tightly than individual parameters, and near-optimal solutions can occupy distinct modes with similar curve errors but substantially different physical parameter values.
The degeneracy in curve space is structured, not random scatter.

\subsubsection{Clustering and real-space structure}
\label{sec:clustering}

The correlation structure and multimodality indicate that near-optimal solutions occupy distinct regions of parameter space.
To identify these regions, we cluster the retained solutions with HDBSCAN and visualize their low-dimensional structure with PCA (Methods, section~\ref{sec:solution_analysis}).

\begin{figure}[b!]
    \centering
    \includegraphics[width=0.99\linewidth]{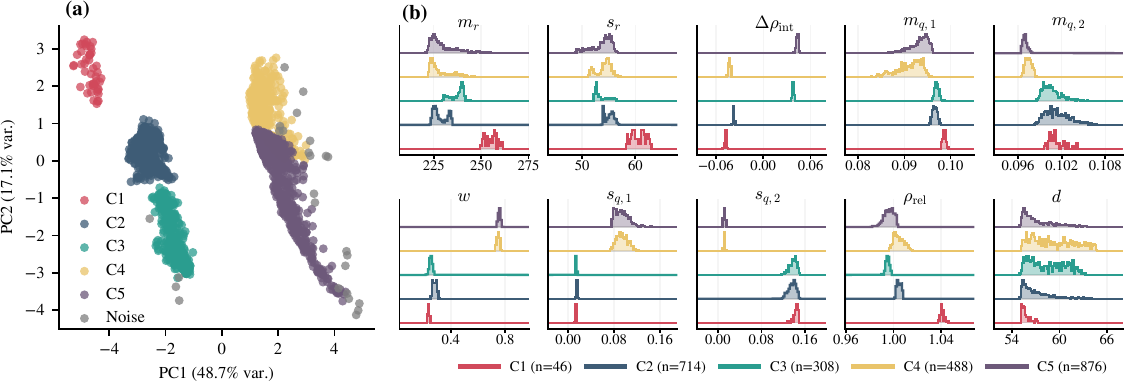}
    \caption{ Structure of the inferred parameter space for retained near-optimal solutions. \textbf{(a)} Retained solutions in parameter space, colored by HDBSCAN cluster assignment (noise in gray). Distinct clusters appear despite nearly indistinguishable SAXS fits. \textbf{(b)} Cluster-resolved marginal parameter distributions, with dashed lines indicating cluster medians.\looseness-1 }
    \label{fig:clustering}
\end{figure}

Figure~\ref{fig:clustering} shows that the retained near-optimal solutions for the MC3 data separate into five well-defined clusters in parameter space, identified by HDBSCAN from the data (the cluster count is not imposed and is unrelated to the number of synthetic benchmarks).
Stratifying by cluster reveals that several parameters show cluster-specific peaks, most clearly for $\Delta\rho_{\mathrm{int}}$, $w$, $s_{q,1}$, $s_{q,2}$, and $\rho_{\mathrm{rel}}$ (Fig.~\ref{fig:clustering}b).
Other parameters, such as the shell thickness $d$, remain broadly distributed within and across clusters, indicating weaker constraints.
The clustering makes the main modes of non-identifiability concrete: the five clusters correspond to coordinated combinations of parameters, not independent scatter in individual parameters.
Each reconstruction shown is the medoid solution for its cluster (see section~\ref{sec:solution_analysis}), a representative snapshot rather than a unique summary.\looseness-1

\begin{figure}[t!]
    \centering
    \includegraphics[width=\linewidth]{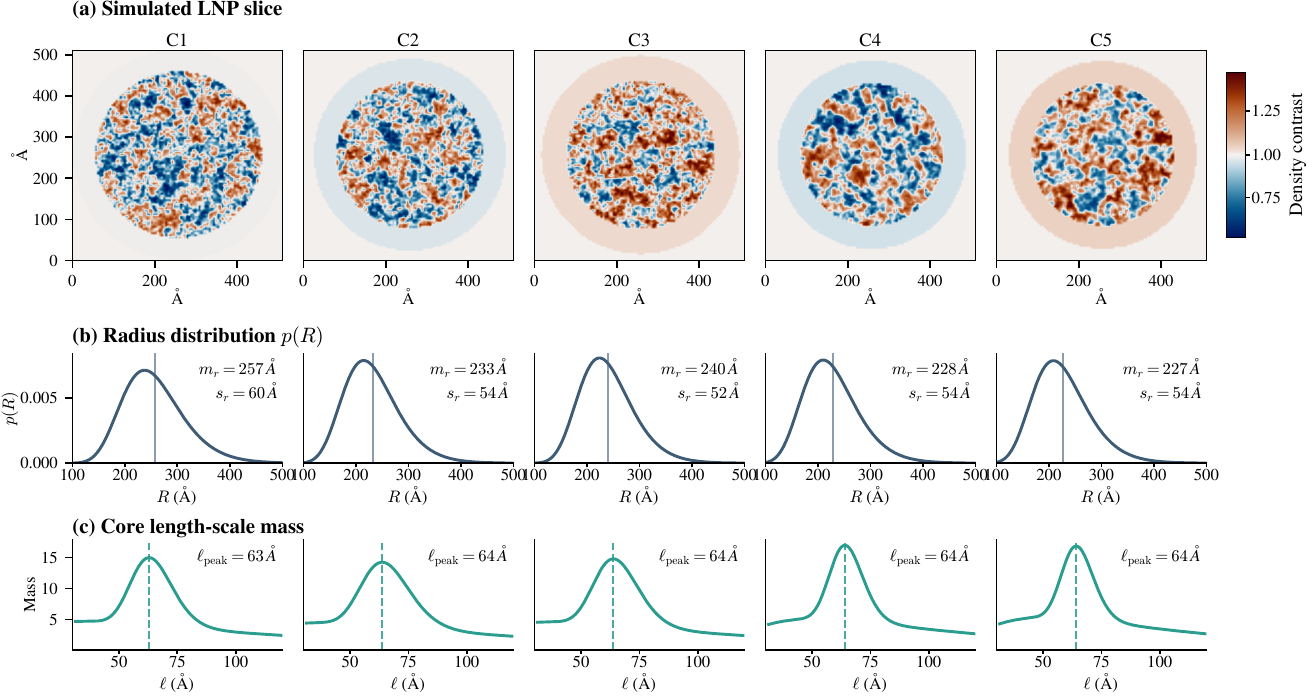}
    \caption{
Representative real-space morphologies, radius polydispersity, and inferred core length-scale mass for each parameter cluster among the retained near-optimal SAXS solutions.
\textbf{(a)} 2D cross-sections through the reconstructed 3D electron-density fields.
\textbf{(b)} Corresponding truncated log-normal radius distributions parameterized by $(\mu,\sigma)$.
\textbf{(c)} Shell-weighted spectral mass implied by the core random-field parameters, plotted against the equivalent length scale $\ell = 2\pi/q$ in \AA, with the dashed line indicating the peak length scale.
Across the three panels, near-equivalent SAXS fits correspond to distinct interior morphologies, particle-size distributions, and characteristic core length scales.
    }
    \label{fig:poly}
\end{figure}

To connect these modes to physical structure, Fig.~\ref{fig:poly} shows representative reconstructions, radius distributions, and core length-scale spectra for each cluster.
The five clusters are organized into two groups by shell-to-interior contrast.
Clusters~C2 and C4 have $\rho_{\mathrm{rel}} < 1$, meaning the shell electron density is lower than the mean interior density.
Clusters~C3 and C5 have $\rho_{\mathrm{rel}} > 1$, placing the shell denser than the interior mean.
Cluster~C1 occupies an intermediate position where $\rho_{\mathrm{rel}} \approx 1$, effectively removing the shell contrast and making the particle appear as a core-only structure with a larger effective electron-density radius, $m_r \approx 257$~\AA (core radius approximately $200$~\AA\ after accounting for the shell thickness $d \approx 56$~\AA). Because SAXS probes the electron-density boundary rather than the hydrodynamic or projected visual boundary, this value is not directly comparable to the DLS and cryo-TEM estimates.

Within each contrast group, the two clusters differ primarily in the distribution of spectral weights of the interior structure.
For C2 and C4, which share $\rho_{\mathrm{rel}} < 1$ and have similar mean radii ($m_r \approx 233$~\AA{} for C2, $m_r \approx 228$~\AA{} for C4), the distinguishing parameters are $w$, $s_{q,1}$, and $s_{q,2}$.
Cluster~C4 concentrates spectral weight on the second component ($w \approx 0.7$, $s_{q,1} \approx 0.1$, $s_{q,2} \approx 0.02$), producing a broad interior length-scale distribution.
Cluster~C2 shifts weight to the first component ($w \approx 0.3$, $s_{q,1} \approx 0.02$, $s_{q,2} \approx 0.14$), producing a similarly broad distribution but through a different spectral decomposition.
The same pattern appears in the higher-contrast group: C5 ($w \approx 0.7$) has a narrower core length-scale peak than C3 ($w \approx 0.3$), mirroring the C4/C2 relationship (Fig.~\ref{fig:poly}c).

The core length scale distributions (Fig.~\ref{fig:poly}c) provide a quantitative picture of what the SAXS data constrain about interior heterogeneity: the peak position is constant, but the width of the spectral mass varies across clusters, showing that the data determine a small range of plausible interior length scales rather than a single value.

The near-optimal ensemble thus contains multiple physically distinct solution modes, all producing essentially indistinguishable SAXS fits, and the data alone do not distinguish among them.
The baseline comparison below (section~\ref{sec:baselines}) examines how the inferred size distribution changes under simpler analytical models.

\subsubsection{Sensitivity analysis}
\label{sec:local_sens}

The ensemble analysis above characterizes non-identifiability globally by surveying the space of near-optimal solutions.
A complementary question is local: around any given solution, which parameters does the SAXS curve respond to most strongly, and which parameter perturbations produce similar (or opposing) changes in the predicted intensity?
Answering this requires the Jacobian of the forward model, which is available analytically because the surrogate and polydispersity layer are fully differentiable.

\begin{figure}[t]
    \centering
    \includegraphics[width=0.999\linewidth]{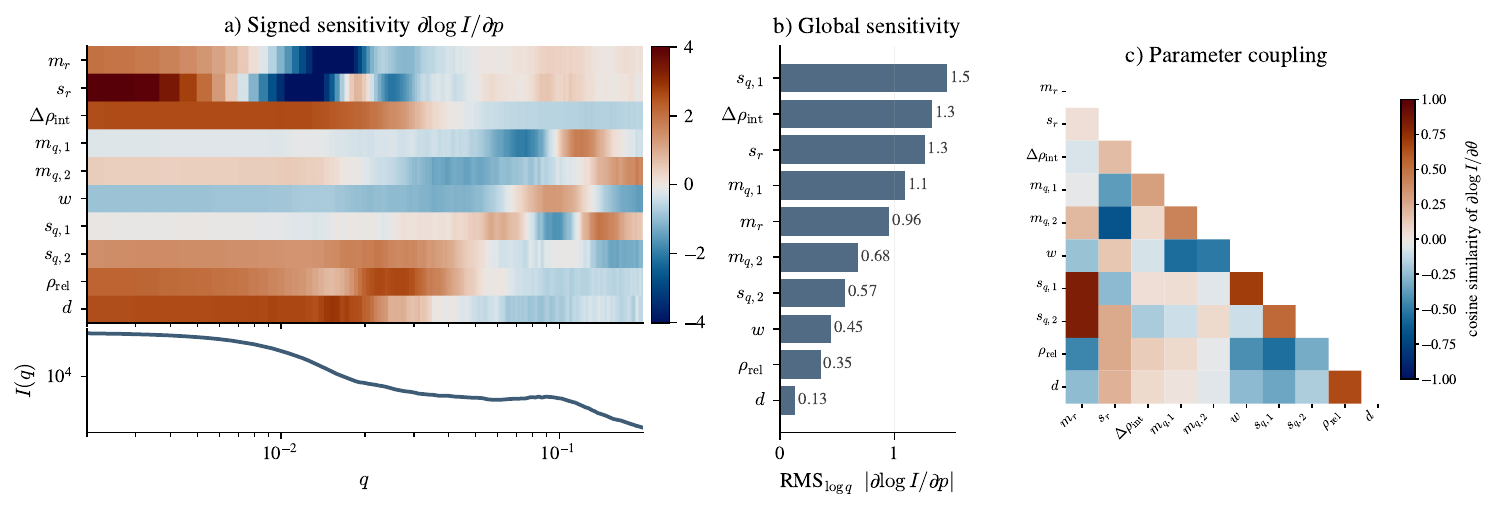}
    \caption{
    Sensitivity and parameter coupling for the differentiable polydisperse SAXS predictor, aggregated over retained near-optimal solutions.
    \textbf{(a)} Median Jacobian of the $q^2$-normalized log-intensity with respect to each fitted parameter, shown as a signed heatmap over $q$ (rows $L_2$-normalized).
    Bottom: corresponding fitted $I(q)$ (log--log). 
    \textbf{(b)} Global sensitivity score per parameter (RMS over $q$ of the Jacobian magnitude).
    \textbf{(c)} Parameter coupling matrix given by cosine similarity between Jacobian profiles over $q$ (positive: similar effects, negative: opposing).
    }
    \label{fig:sensitivity_signed}
\end{figure}

For each retained solution, we compute the Jacobian of the $q^2$-normalized log-intensity curve with respect to the fitted parameters (Methods, section~\ref{sec:solution_analysis}).
Figure~\ref{fig:sensitivity_signed} summarizes the median Jacobian profiles across the ensemble, a global per-parameter sensitivity score, and a coupling matrix based on cosine similarity of Jacobian columns.

\paragraph{Scale separation in the Jacobian.}
The Jacobian heatmap (Fig.~\ref{fig:sensitivity_signed}a) reveals a clear separation between parameters that primarily affect different $q$ regions.
The size-distribution parameters $m_r$ and $s_r$ dominate at low $q$ (below approximately $0.02$~\AA$^{-1}$), consistent with particle sizes determining large-length-scale scattering.
Their signed sensitivities change sign across $q$, reflecting that size perturbations shift the locations and relative amplitudes of oscillatory features in the log-intensity profile.
The shell thickness $d$ also shows a localized contribution at low $q$, with a peak near $q \approx 2\pi/d$.
Interior-structure parameters ($\Delta\rho_{\mathrm{int}}$, $m_{q,1}$, $s_{q,1}$, $s_{q,2}$) and the relative shell contrast $\rho_{\mathrm{rel}}$ contribute primarily at intermediate and high $q$, where interior heterogeneity and the shell's contribution to the scattering are probed. The spectral weight $w$ and the second spectral width $s_{q,2}$ show broader $q$-range contributions than $m_{q,1}$ and $s_{q,1}$. The weight $w$ redistributes spectral mass between the two log-normal components, affecting the curve wherever either component has support. The width $s_{q,2}$ controls the breadth of the second component, which can extend its influence into the low-$q$ regime depending on the component's location.

\paragraph{Global sensitivity.}
The global sensitivity score (Fig.~\ref{fig:sensitivity_signed}b) summarizes, for each parameter, how much the predicted curve changes on average across all $q$ when that parameter is perturbed by a small amount.
The GRF spectrum width $s_{q,1}$ attains the highest median score.
This is consistent with its role in controlling the breadth of the interior length-scale distribution: because $s_{q,1}$ affects the shape of the power spectrum over a wide range of $q$, even a small perturbation induces changes across much of the curve.

The interior density offset $\Delta\rho_{\mathrm{int}}$ has the second-highest score. Because SAXS intensity depends on the squared electron-density contrast, shifting $\Delta\rho_{\mathrm{int}}$ rescales the entire interior scattering contribution, so a small perturbation produces a broad change in $I(q)$ across the $q$-range where the interior contributes.
The size parameters $s_r$ and $m_r$ follow, with their influence concentrated at low $q$ but strong enough there to produce high average scores.
The shell parameters $\rho_{\mathrm{rel}}$ and $d$ have lower global sensitivity, indicating that near the retained solutions they induce comparatively smaller or more localized curve changes.

A high global sensitivity does not, by itself, mean that a parameter is well constrained by the data.
A parameter with high sensitivity but strong coupling to another parameter may still be poorly identifiable because co-varying the two can leave the curve nearly unchanged.
The coupling matrix addresses this.

\paragraph{Parameter coupling.}
The coupling matrix (Fig.~\ref{fig:sensitivity_signed}c) measures cosine similarity between the Jacobian profiles of parameter pairs.
High positive values indicate that two parameters induce similar $q$-dependent changes in the curve, so the data cannot easily distinguish perturbations in one from perturbations in the other.
High negative values indicate opposing effects that can cancel when the two parameters co-vary.

The strongest positive couplings are between $s_{q,1}$ and $m_r$, and between $s_{q,2}$ and $m_r$.
This means that changes in the interior spectral widths produce curve modifications that resemble changes in mean particle size, providing a local, mechanistic explanation for the size-vs-interior trade-offs identified in the ensemble correlation analysis (section~\ref{sec:param_tradeoffs}).
The strongest negative coupling is between $m_{q,2}$ and $s_r$: shifting the position of the second spectral peak and adjusting the radius spread have opposing effects on the curve, so co-varying them in opposite directions can preserve the fitted intensity.

These coupling patterns explain, at the level of the forward model's local geometry, why many distinct parameter settings yield nearly indistinguishable curves.
The data tightly constrain certain combinations of parameters, but individual parameters can trade off against each other along the directions identified here.
This local analysis is consistent with the global ensemble structure reported in Sections~\ref{sec:param_tradeoffs}--\ref{sec:clustering}, and it identifies specific compensatory directions that complementary measurements (e.g., contrast-variation SAXS or SANS, or independent size constraints) would need to break.

\subsection{Comparison with analytical baseline models}
\label{sec:baselines}

The heterogeneous core--shell model used in this paper has more parameters and a more complex forward map than standard analytical form-factor models.
This comparison asks two questions.
First, at what $q$-range does the full model become necessary for accurate curve fitting?
Second, what happens to the inferred size-distribution parameters when the model lacks the physics to represent interior structure?

We fit two analytical baseline models to the same experimental SAXS profile: a polydisperse homogeneous sphere (two free parameters: mean radius $\bar{R}$ and standard deviation $\sigma_R$) and a polydisperse core--shell sphere (four free parameters: $\bar{R}$, $\sigma_R$, shell thickness $d$, and contrast ratio $c$).
Full details of the form factors, loss functions, and fitting procedures are given in Supplementary section~S8.
All three models use the same truncated log-normal radius distribution, the same Gauss--Legendre quadrature for polydispersity integration, the same multi-start gradient-based optimizer (AdamW with automatic differentiation in PyTorch), and the same composite fitting loss.
Any difference in fit quality or inferred parameters, therefore, reflects the forward model's expressiveness rather than the optimizer or fitting procedure.

Each model is fitted at three $q$ cutoffs ($q_{\max} \in \{0.02, 0.05, 0.10\}$~\AA$^{-1}$) to show which structural features become accessible as more of the curve is included.
Full curves over the complete measured $q$-range are stored for every restart, regardless of the fitting window, to enable post-hoc evaluation.

\begin{figure}[t!]
    \centering
    \includegraphics[width=0.85\linewidth]{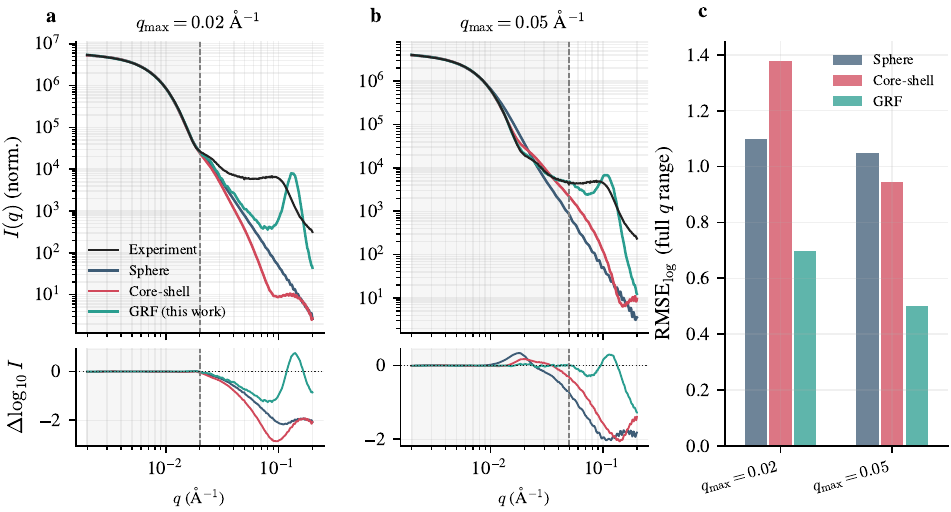}
    \caption{
    \textbf{Comparison of analytical baseline models with the heterogeneous core--shell model.}
    \textbf{(a)--(b)}~Best-fit curves (top) and log-intensity residuals (bottom)
    for each fitting cutoff ($q_{\max} = 0.02$, $0.05$~\AA$^{-1}$;
    shaded band).
    All curves are shown over the full measured $q$ range.
    At low $q$, all three models fit comparably; at higher cutoffs, the sphere
    and core--shell models show systematic residuals where shell and interior
    features contribute, while the heterogeneous core--shell model remains accurate.
    \textbf{(c)}~$\mathrm{RMSE}_{\log}$ evaluated over the complete $q$ range
    for each model and cutoff.
    The analytical models degrade as the fitting range extends; the heterogeneous core--shell model maintains low error throughout.
    Full details in Supplementary section~S8.
    }
    \label{fig:comparison_main}
\end{figure}

\paragraph{Curve fit quality across \texorpdfstring{$q$}{q} cutoffs.}
Figure~\ref{fig:comparison_main}a--b shows the best-fit curves at all three cutoffs alongside log-intensity residuals.
At $q_{\max} = 0.02$~\AA$^{-1}$, all three models reproduce the experimental curve comparably, since only particle size is probed at low $q$.
As the fitting range extends, the sphere and core--shell models develop systematic residuals in the regimes where shell modulation and interior heterogeneity contribute, while the heterogeneous core--shell model remains accurate across the full $q$-range.
Figure~\ref{fig:comparison_main}c quantifies this: full-range $\mathrm{RMSE}_{\log}$ degrades with increasing $q_{\max}$ for the analytical models but remains low for the heterogeneous model.

\paragraph{Size-distribution bias in simpler models.}
The simpler spherical and core--shell analytical models do not capture all scattering features within the fitted $q$ window because they lack degrees of freedom to account for interior heterogeneity.
This missing flexibility distorts their inferred size-distribution parameters (Supplementary Fig.~S7).
An analytical core--shell model augmented with an explicit Gaussian correlation peak can reproduce the incoherent scattering feature for a single parameter setting (Supplementary Fig.~S9), but this approach requires manual tuning and does not support the systematic ensemble analysis enabled by the GRF parameterization.
  
In contrast, the heterogeneous core--shell model directly incorporates the interior contribution via the GRF spectral parameters, so the inferred radius distribution does not need to compensate for scattering features that simpler models omit.

A subtler effect is visible already at low $q$, where all three models fit the curve comparably.
The range of mean radii across near-optimal solutions differs between models: the sphere model accepts a narrow band of mean radii (approximately $\bar{R} \in [227, 233]$~\AA), the core--shell model accepts a somewhat wider range ($\bar{R} \in [225, 242]$~\AA), and the heterogeneous core--shell model accepts the widest range ($\bar{R} \in [210, 245]$~\AA) (Supplementary Fig.~S7).
This happens because interior heterogeneity, even at low $q$, slightly modulates the scattering.
The two simpler models, which lack interior degrees of freedom, narrow their range of acceptable size parameters, but that apparent precision is an artifact of an oversimplified forward model.
A model that cannot represent interior structure rules out size-distribution solutions that a more complete model retains, potentially excluding the true parameter values and underestimating the real uncertainty.

\paragraph{When is the full model needed?}
This comparison provides direct evidence for the role of interior heterogeneity. When the fitting range is restricted to $q_{\max} \leq 0.02$~\AA$^{-1}$, only particle size is probed, and the analytical models are adequate for curve fitting, though they already underestimate the range of plausible size-distribution parameters. Once the fitting range extends to $q_{\max} \geq 0.05$~\AA$^{-1}$, so that shell and interior features enter the fit, the analytical models introduce systematic errors in the inferred size distribution that can be detected by comparison with cryo-TEM and DLS. The heterogeneous core--shell model is therefore needed whenever the fitting range extends beyond $q_{\max} \approx 0.02$~\AA$^{-1}$.  More elaborate analytical parameterizations, such as multi-shell models \cite{li2025mesoscopic}, introduce additional interior degrees of freedom but remain constrained to prescribed geometries.

Even when the fitting range is restricted to low $q$, the heterogeneous model still gives a broader and more realistic characterization of size-distribution uncertainty than the analytical baselines.

\subsection{Surrogate accuracy and computational cost}
\label{sec:surrogate_performance}

All preceding analyses depend on the neural surrogate accurately and rapidly approximating the realization-averaged forward map.
The surrogate replaces the full numerical pipeline (3D GRF generation, Fourier-slice SAXS computation, and averaging over $n=50$ realizations) with a single neural-network forward pass that maps structural parameters directly to a log-intensity profile.
This section validates that the approximation is accurate enough to support the identifiability analyses above and quantifies the speedup that makes them feasible.\looseness-1

\begin{figure}[t!]
    \centering
    \includegraphics[width=0.98\linewidth]{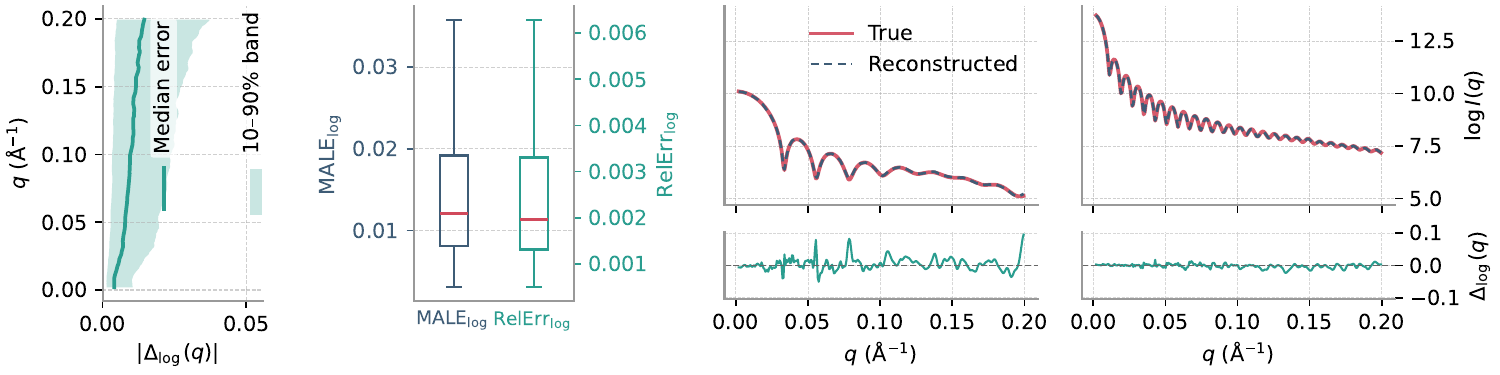}
    \caption{
    Validation of the DCT-based surrogate on a held-out test set of 1024 simulated SAXS curves.
    \textbf{Left:} Pointwise log-error $|\Delta_{\log}(q)|$ with median and 10--90th percentile band.
    \textbf{Middle:} Test-set distributions of $\mathrm{MALE}_{\log}$ and $\mathrm{RelErr}_{\log}$.
    \textbf{Right:} Representative predicted and reference profiles with signed residuals $\Delta_{\log}(q)$.
    }
    \label{fig:predictor_results}
\end{figure}

\paragraph{Surrogate accuracy.}
We evaluate the DCT-based surrogate on a held-out test set of 1024 simulated SAXS curves (Methods, section~\ref{sec:eval_metrics}).
Across the test set, the surrogate achieves $\mathrm{MALE}_{\log}=0.028 \pm 0.006$ (mean$\pm$std), and the median $\mathrm{RelErr}_{\log}$ is below $0.2\%$.
Pointwise errors remain small across the full $q$-range, with only modest variation between low-$q$ and high-$q$ regions (Fig.~\ref{fig:predictor_results}, left).
Signed residuals are centered close to zero (Fig.~\ref{fig:predictor_results}, right), with no systematic bias.
The surrogate preserves both the global decay shape and the higher-$q$ oscillatory structure relevant to fitting interior heterogeneity.

To confirm that the surrogate error does not drive the ensemble structure reported in Sections~\ref{sec:param_tradeoffs}--\ref{sec:clustering}, we note that the surrogate approximation error ($\mathrm{MALE}_{\log} \approx 0.028$) is substantially smaller than the parameter-space spread within the retained ensembles (see Supplementary Section~S5 for accuracy across the full parameter domain).
The organized structure of the near-optimal solutions (multimodal marginals, strong inter-parameter correlations, five distinct clusters) cannot be produced by random surrogate noise.
The same trade-off patterns appear in the synthetic benchmarks, where the surrogate is evaluated against known ground truth, confirming that structured non-identifiability is a property of the forward model rather than an artifact of approximation error. In addition, the polydisperse forward evaluation averages the surrogate output over $K=64$ radii, reducing the effect of any pointwise surrogate error on the fitted curve. 

\begin{figure}[t!]
    \centering
    \includegraphics[width=0.97\linewidth]{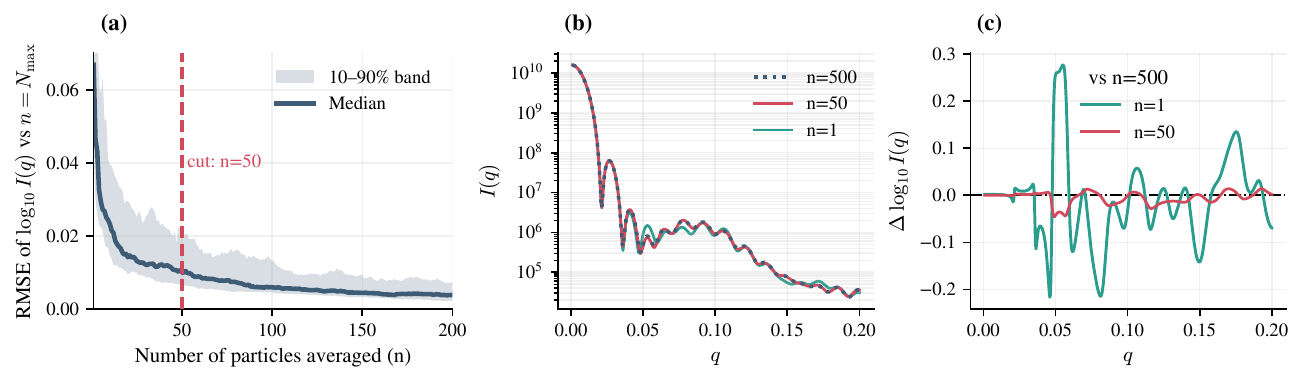}
    \caption{
    Convergence of GRF-averaged SAXS curves at fixed parameters.
    (a) Over 20 random parameter settings: median and 10--90th percentile RMSE in $\log_{10} I(q)$ between the running mean at $n$ realizations and a reference ($n=N_{\max}$); the vertical line indicates $n=50$.
    (b) Example curves for $n=1$, $n=50$, and $n=N_{\max}$.
    (c) Residuals in $\log_{10} I(q)$ versus the $n=N_{\max}$ reference.
    }
    \label{fig:fig_convergence_triptych}
\end{figure}

\paragraph{GRF averaging convergence.}
Because the interior density is generated by a stochastic model, monodisperse SAXS curves exhibit random variability when computed from a finite number of GRF realizations.
Figure~\ref{fig:fig_convergence_triptych}(a) shows that the RMSE between the running-mean curve at $n$ realizations and a high-sample reference drops rapidly with $n$, with clear diminishing returns beyond $n \approx 50$.
At $n=50$, the median RMSE is already ${\sim}10^{-2}$ in $\log_{10} I(q)$, and residual Monte Carlo variability is small compared to the fitting tolerances used to define the near-optimal ensembles.
We therefore use $n=50$ realizations per parameter setting for generating stable training targets.

\paragraph{Computational cost.}
Table~\ref{tab:runtime_breakdown} compares the stochastic simulator with the surrogate-based forward model. The surrogate reduces the cost of a single polydisperse forward evaluation (with $K=64$ quadrature radii) from approximately 49~s (including averaging over $n=50$ GRF realizations per radius) to 2.4~ms, a speedup of roughly four orders of magnitude. This reduction is what makes multi-start gradient-based fitting practical: 128 restarts with 1000 iterations each complete in under 40~s on a single GPU, whereas embedding the stochastic simulator in the same optimization loop would require repeated GRF generation and averaging at each step, which is not computationally feasible. The 8000 restarts used for the experimental analysis scale linearly from this baseline.

\begin{table}[ht]
\centering
\caption{
Runtime comparison on a single GPU, excluding I/O and JIT compilation.
Simulator timings use $n=50$ GRF realizations per monodisperse curve; polydisperse evaluation uses $K=64$ quadrature radii.
}
\label{tab:runtime_breakdown}

\footnotesize
\setlength{\tabcolsep}{6pt}
\renewcommand{\arraystretch}{1.12}

\begin{tabular}{@{}p{6.5cm} r r@{}}
\toprule
\textbf{Task} & \textbf{Simulator} & \textbf{Surrogate} \\
\midrule
Monodisperse forward, one $(r,\theta)$
& $1.53 \pm 0.14$ s
& $1.26 \pm 0.00$ ms \\

Monodisperse forward, batch size 512
& $13.1 \pm 1.2$ min
& $1.4 \pm 0.1$ ms \\

Polydisperse forward, one $(\theta,\xi)$ with $K=64$
& $\approx 49.0 \pm 4.6$ s
& $2.43 \pm 0.20$ ms \\

One restart, 1000 iterations, forward and backward
& n/a
& 11.8 s \\

Multi-start, 128 restarts
& n/a
& 37.1 s \\
\bottomrule
\end{tabular}
\end{table}

\section{Discussion}\label{sec:discussion}
We have developed a machine learning--accelerated, differentiable inference framework for SAXS analysis of heterogeneous, polydisperse lipid nanoparticles and used it to ask what a measured scattering profile can and cannot jointly determine about LNP structure and size distribution.
The synthetic benchmarks and the experimental MC3 LNP analysis lead to the same conclusion: close agreement in curve space does not imply unique parameter recovery.
Near-optimal solutions form structured clusters in parameter space even when the predicted $I(q)$ curves are essentially indistinguishable, and the dominant source of ambiguity is compensation between size-distribution parameters and the GRF spectral parameters that control interior heterogeneity.

Reporting a single best-fit parameter set, therefore, understates the ambiguity inherent in the measurement. The
near-optimal ensemble is the more informative output, identifying which parameters are stable across fits and which
remain ambiguous. In the MC3 data, overall particle size and the broad features of the scattering curve are consistently constrained, while finer structural details are not.
Analytical models without interior degrees of freedom, when fitted over $q$ ranges that probe interior structure, compensate by shifting the inferred size distribution away from the values obtained at low $q$, whereas the heterogeneous core--shell model incorporates the interior contribution via the GRF spectral parameters without distorting the radius distribution.
The organized cluster structure is not driven by surrogate approximation error, which is more than an order of magnitude smaller than the parameter-space spread (section~\ref{sec:surrogate_performance}).

Beyond exposing this structure, the ensemble analysis itself indicates where additional information would be most valuable: the dominant trade-off involves size-distribution parameters, so independent size estimates are the most direct way to narrow the retained ensemble.
For the MC3 LNP system studied here, the retained near-optimal solutions infer mean radii systematically below both the DLS value of $323 \pm 79$~\AA\ and the cryo-TEM estimate of $346 \pm 62$~\AA, and even Cluster~C1, the largest cluster at $m_r \approx 257$~\AA\ (Sec.~\ref{sec:clustering}), remains well below these external values.

Some offset is expected because SAXS is sensitive to the electron-density contrast boundary, whereas DLS provides hydrodynamic radii that generally overestimate the true size, and cryo-TEM images LNPs in a vitrified state that may not be representative of the solution. The resulting differences in estimated size across the three techniques are therefore not surprising, and the methods are broadly consistent with one another. This also means, however, that DLS and cryo-TEM offer little power to discriminate among the near-optimal clusters.
A more targeted approach would be contrast-variation SAXS or small-angle neutron scattering (SANS), which could resolve interior-structure modes directly, as these techniques manipulate the scattering contrast of specific components rather than the overall size.

A second structural route is to use a more parsimonious description of interior heterogeneity, but in practice, we could not reduce the parameterization below the two-component spectral form used here without a substantial loss in goodness of fit.

The identifiability conclusions reported here are conditional on the modeling choices.
The forward model assumes spherical geometry, a homogeneous shell, and a stochastic interior described by a GRF, which serves as a flexible statistical model of interior heterogeneity rather than a mechanistic account of lipid self-assembly.
Although the GRF is, in principle, a highly flexible interior model, the non-identifiability we report already appears under the restrictive two-component spectral parameterization used throughout, so it is not simply an artifact of model overflexibility.
Alternative structural models, non-spherical geometries, or extended $q$-ranges may change which parameter directions are effectively constrained. The experimental results are limited to a single MC3 LNP formulation, so the specific degeneracy structure should be characterized independently for each new system.
A natural methodological extension is to replace multi-start optimization with Bayesian inference, thereby adding posterior mode weights and marginal credible intervals to the ensemble analysis used here.
For the MC3 system, and for other heterogeneous, polydisperse nanoparticles, including those probed by SANS, the framework identifies which parameter directions remain degenerate and which complementary measurements would resolve them, making explicit what the data can and cannot determine about structure and size distribution.

\section{Methods}\label{sec:methods}

We first describe the experimental setup used for LNP formulation and data acquisition, and then present the computational modeling framework for SAXS-based structural inference. 

\subsection{Experimental methods}\label{sec:experimental}

\paragraph{LNP formulation.}
Lipid nanoparticles (LNPs) were prepared by rapid microfluidic mixing of a lipid solution in ethanol with an aqueous mRNA solution in RNase-free citrate buffer (50 mM, pH 3.0). The payload was 90\% eGFP mRNA (L-7201, TriLink) and 10\% Cy5-labeled eGFP mRNA (L-7701, TriLink). Lipids (DLin-MC3-DMA/cholesterol/DSPC/DMPE-PEG, 50:38.5:10:1.5 mol/mol) were dissolved in 99.5\% ethanol at 12.5 mM and mixed on a NanoAssemblr Ignite at a 1:3 lipid:aqueous flow-rate ratio (total 12 mL min$^{-1}$), yielding a lipid:mRNA mass ratio of approximately 10:1. Post-mixing, LNPs were dialyzed overnight against PBS (pH 7.4; Slide-A-Lyzer, 10 kDa MWCO), then dialyzed a second time into a freezing buffer comprising 20 mM Tris (pH 7.5) with 8\% sucrose. LNPs were concentrated (Amicon, 30 kDa MWCO) and sterile-filtered (0.2 $\mu$m). Encapsulation efficiency was 98\%, quantified by RiboGreen after Triton X-100/TE disruption of LNPs.\looseness-1

\paragraph{SAXS measurements.}
Small-angle X-ray scattering (SAXS) experiments were carried out at the CoSAXS beamline at MAX IV (Lund, Sweden). LNP suspensions were delivered to a temperature-controlled flow-through quartz capillary (1.0 mm outer diameter) using an automated BioCube sampler (Xenocs). All measurements were performed at 25 $^{\circ}$C. Scattering data were collected with a dual-detector setup comprising EIGER2 4M and PILATUS3 2M detectors (Dectris AG) at an X-ray wavelength of 1.0 \AA. Merged datasets from the two detectors provided a scattering vector $q$ range of 0.001--1.9 \AA$^{-1}$. Intensities were corrected for transmission and placed on an absolute scale using water as the primary standard, and sample profiles were background-subtracted using the corresponding buffer measurements.
The analyses in this work use a restricted range of $q = 0.001$--$0.200$~\AA$^{-1}$, which covers the length scales described by the forward model (overall particle size, shell modulation, and interior heterogeneity). Higher-$q$ features, which probe molecular-scale structure not represented in the model, are outside the analysis window.

\paragraph{Cryogenic transmission electron microscopy.}
Cryogenic transmission electron microscopy (cryo-TEM) was performed on a JEOL JEM-2100Plus operated at 200 kV with a lanthanum hexaboride electron source. Micrographs were acquired using a TVIPS TemCam-XF416 under low-dose conditions optimized for low contrast. Lacey-carbon support grids were prepared in an environmental chamber set to 15 $^{\circ}$C and 95\% relative humidity. A 3 $\mu$L aliquot of sample was applied to each grid and blotted for 2 s to remove excess liquid, after which the grids were plunged into liquid ethane to vitrify the thin aqueous film and preserve native structure in amorphous ice. Vitrified grids were transferred to the cryogenic holder under continuous cryogenic conditions and inserted into the microscope. During imaging, the specimen was maintained below -160 $^{\circ}$C, with data collection at approximately -178 $^{\circ}$C to prevent devitrification and ensure stability.

The images have pixel size $\sim$ 2.3 \AA\ and resolution 4096x4096 pixels (FOV $\sim$ 0.95 $\mu$m). Particle sizes were extracted using manual annotation of the contours of 400 particles. Each area inside a contour was converted to an equivalent radius. 

\paragraph{Dynamic light scattering.}
Dynamic light scattering (DLS) measurement for particle size characterization was performed on a Zetasizer Nano ZS (Malvern Panalytical). The LNPs were diluted in phosphate-buffered saline (PBS, pH 7.4). Measurements were performed at 25 $^{\circ}$C using a fixed backscattering angle of 173$^{\circ}$ with a 632.8 nm He--Ne laser. The number-weighted distribution of (hydrodynamic) radii was modeled as a Schultz distribution, including corrections based on Mie scattering theory, yielding a model for the field autocorrelation function $g_1(\tau)$. Actual fitting was performed to the measured intensity-intensity autocorrelation $g_2(\tau)$. Measurements were performed in triplicate. 
Raw autocorrelation fits, and the corresponding Schultz size distributions are shown in Supplementary Section~S9 (Fig.~S11).

\subsection{Computational pipeline}\label{sec:computational_pipeline}

\begin{figure}[ht]
    \centering
    \includegraphics[width=0.999\linewidth]{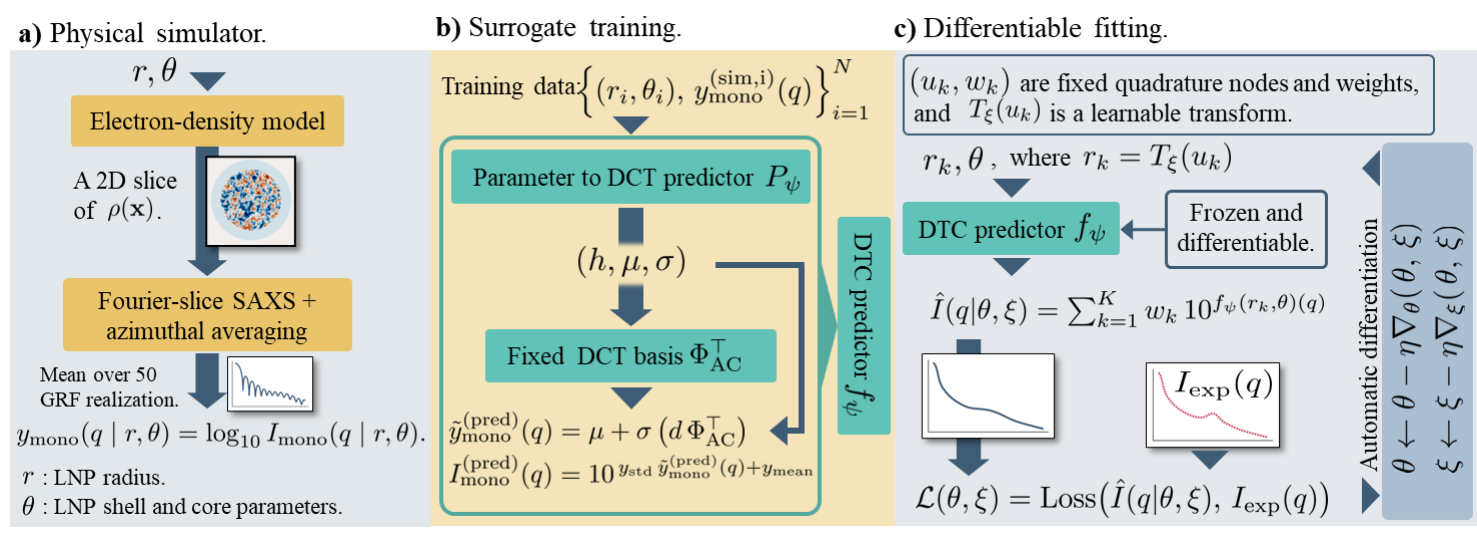}
\caption{
\textbf{Detailed computational workflow corresponding to the conceptual overview in Fig.~\ref{fig:model_overview}.}
\textbf{(a) Physical simulation.} From \((r,\theta)\), we generate an electron-density field \(\rho(\mathbf{x})\) and compute the monodisperse SAXS curve via Fourier-slice SAXS and azimuthal averaging. Curves are averaged over 50 independent GRF realizations.
\textbf{(b) Surrogate model.} A residual MLP \(P_\psi\) is trained on simulated data to predict DCT-parameterized log-intensity curves using a fixed DCT basis.
\textbf{(c) Differentiable fitting.} The pretrained surrogate is combined with a differentiable polydispersity model and fitted to experimental SAXS data by gradient-based optimization with automatic differentiation.
}
\label{fig:saxs_framework}
\end{figure}

Inferring structural parameters from a polydisperse SAXS profile requires a forward model that captures interior heterogeneity, a fast approximation that makes repeated evaluation practical, and a differentiable fitting layer that can optimize both structural and size-distribution parameters jointly (Fig.~\ref{fig:saxs_framework}).

Additional details on training-data diversity, network architecture ablation, surrogate robustness across the parameter domain, polydispersity quadrature convergence, and the PyTorch implementation are provided in the Supplementary Information.

\subsubsection{Physical simulation model and SAXS computation}\label{sec:saxs_simulations}

Each LNP is modeled as a sphere of radius \(r\) comprising a uniform shell of thickness \(d\) surrounding a disordered core.
The structural parameters, collectively denoted by
\[
\theta = (\theta_{\mathrm{core}},\,\theta_{\mathrm{shell}}),
\]
encode both the interior core structure and the shell geometry.
The core parameters \(\theta_{\mathrm{core}}\) describe the Gaussian random field (GRF) model that generates spatial density fluctuations,
including a two-component spectral model and a logistic mapping used to obtain a bounded interior density field.
The shell parameters \(\theta_{\mathrm{shell}}\) specify the shell thickness \(d\) and a relative shell contrast parameter (see below).

\paragraph{Electron-density synthesis.}
The disordered lipid core is represented by an isotropic GRF \(G(\mathbf{x})\).
We generate \(G\) by filtering Fourier-domain white noise \(\widetilde{W}(\mathbf{q})=\mathcal{F}\{W(\mathbf{x})\}\) with a radial spectrum \(\sqrt{P(q)}\) and inverse-transforming,
\[
G(\mathbf{x}) = \mathcal{F}^{-1}\!\left[\widetilde{W}(\mathbf{q})\,\sqrt{P(q)}\right].
\]
We parameterize the power spectrum \(P(q)\) as a convex mixture of two log-normal components,
\[
P(q)=
(1-w)\,\mathrm{LN}(q; m_{q,1}, s_{q,1})
+
w\,\mathrm{LN}(q; m_{q,2}, s_{q,2}),
\qquad
w \in [0,1], m_{q,1} \leq m_{q,2},
\]
where \(\mathrm{LN}(q;m,\sigma)\) denotes the log-normal density obtained by letting \(\log q \sim \mathcal{N}(m,\sigma^2)\).
The spectrum \(P(q)\) specifies the two-point statistics of the Gaussian random field and sets the characteristic correlation length scales of the interior heterogeneity. \(P(q)\) should be viewed as an interpretable model over interior length scales rather than a mechanistic model of lipid interior structure. For all simulations, we generate fields on a cubic grid with \(N=512\) and spatial resolution \(\Delta x = 5.0~\text{\AA}\).

To obtain a bounded interior density field, we standardize the GRF to zero mean and unit variance, $G_z(\mathbf{x})=(G(\mathbf{x})-\langle G\rangle)/\mathrm{std}(G)$, apply light Gaussian smoothing, and pass the result through a logistic mapping. We then subtract the spatial mean of the mapped field so that the remaining fluctuations have zero mean, and introduce a global offset $\Delta\rho_{\mathrm{int}}$ that controls the mean interior density, and hence the overall contrast relative to the solvent, without altering the correlation structure imposed by $P(q)$:
\begin{equation}
\rho_{\mathrm{int}}(\mathbf{x})
=
1 + \Delta\rho_{\mathrm{int}}
+
\Big(\mathrm{sigmoid}\!\left(G_z(\mathbf{x})\right)-\big\langle\mathrm{sigmoid}(G_z)\big\rangle\Big),
\label{eq:rho_int_def}
\end{equation}
where \(\mathrm{sigmoid}(x) = (1+e^{-x})^{-1}\) is the logistic function and \(\langle\cdot\rangle\) denotes spatial averaging over the grid. The standardization of $G$ (step 1) controls the effective dynamic range of the logistic mapping, while the subtraction of $\langle\mathrm{sigmoid}(G_z)\rangle$ (step 2) ensures that the interior fluctuations have zero spatial mean, so that $\Delta\rho_{\mathrm{int}}$ alone sets the mean interior contrast. 

Throughout, \(\rho(\mathbf{x})\) denotes a dimensionless relative electron-density field normalized such that the solvent has \(\rho=1\).

The full electron-density model is
\[
\rho(\mathbf{x}) =
\begin{cases}
\rho_{\mathrm{int}}(\mathbf{x}), & |\mathbf{x}| \le r-d,\\[3pt]
\rho_{\mathrm{shell}}, & r-d < |\mathbf{x}| \le r,\\[3pt]
1, & |\mathbf{x}| > r,
\end{cases}
\]
with particle radius \(r\) and shell thickness \(d\).
In our implementation, the shell density is tied to the mean interior density via
\(
\rho_{\mathrm{shell}}=\rho_{\mathrm{rel}}\,\langle\rho_{\mathrm{int}}\rangle,
\)
where \(\rho_{\mathrm{rel}}\) is a relative shell-contrast parameter.

\paragraph{Fourier-slice SAXS computation.}

The small-angle X-ray scattering intensity is computed from the electron-density contrast $\Delta\rho(\mathbf{x}) = \rho(\mathbf{x}) - 1$ as
\begin{equation}
I(\mathbf{q}) = \big|\mathcal{F}\{\Delta\rho(\mathbf{x})\}\big|^2 .
\end{equation}

Following \cite{schmidt2007simulation,roding2022}, we approximate the orientation-averaged scattering using a projection-based Fourier-slice approach.
We first integrate along one axis to obtain
\[
\Delta\rho_{\mathrm{proj}}(x,y)
=
\int \Delta\rho(x,y,z)\,dz .
\]
The 2D Fourier transform of this projection corresponds to a central slice of the 3D transform, and we compute\looseness-1
\[
I_{2\mathrm{D}}(q_x,q_y)
=
\big|\mathcal{F}_{2\mathrm{D}}\{\Delta\rho_{\mathrm{proj}}\}\big|^2 .
\]

The particle is embedded in a finite solvent-filled computational domain such that $\Delta\rho=0$ outside the particle.
The projected plane has size $N_f \times N_f$ with $N_f = N + 2p$, where $p$ denotes the solvent margin per side.\looseness-1

The isotropic intensity is obtained by azimuthal averaging,
\[
I_{\mathrm{mono}}(q)
\approx
\left\langle I_{2\mathrm{D}}(q_x,q_y)\right\rangle_{|(qx,qy)|=q}.
\]

The solvent margin was chosen sufficiently large so that simulated SAXS curves are highly resolved over the relevant $q$-range.

\paragraph{GRF sampling and Monte Carlo averaging.}

For a fixed parameter setting $(r,\theta)$, the interior electron-density field is stochastic due to the GRF realization. We therefore approximate the monodisperse scattering curve by averaging over $n$ independent GRF samples:
\[
I_{\mathrm{mono}}(q \mid r,\theta) \;=\; \frac{1}{n}\sum_{\ell=1}^{n} I^{(\ell)}(q \mid r,\theta),
\]
where $I^{(\ell)}$ denotes the curve computed from the $\ell$-th GRF realization using the Fourier-slice procedure described above. We use $n=50$ realizations per parameter setting to obtain stable training targets. Convergence with respect to $n$ is evaluated in Results (section~\ref{sec:surrogate_performance}).

\paragraph{Sampling and normalization.}
The isotropic intensity \(I_{\mathrm{mono}}(q \mid r,\theta)\) is evaluated on an equidistant grid
\(q_i \in [0.001,\,0.200]~\text{\AA}^{-1}\) with \(N_q = 300\).
All simulated curves are stored in log-intensity form,
\begin{equation}
y(q)=\log_{10} I_{\mathrm{mono}}(q \mid r,\theta),
\end{equation}
which compresses the dynamic range, emphasizes weak oscillations at high \(q\), and stabilizes subsequent learning. All key simulation and discretization settings used to generate the training curves are summarized in Table~\ref{tab:sim_settings}.\looseness-1

\paragraph{Model SAXS curves and parameter sensitivity.}

\begin{figure}[t!]
    \centering
    \includegraphics[width=0.7\linewidth]{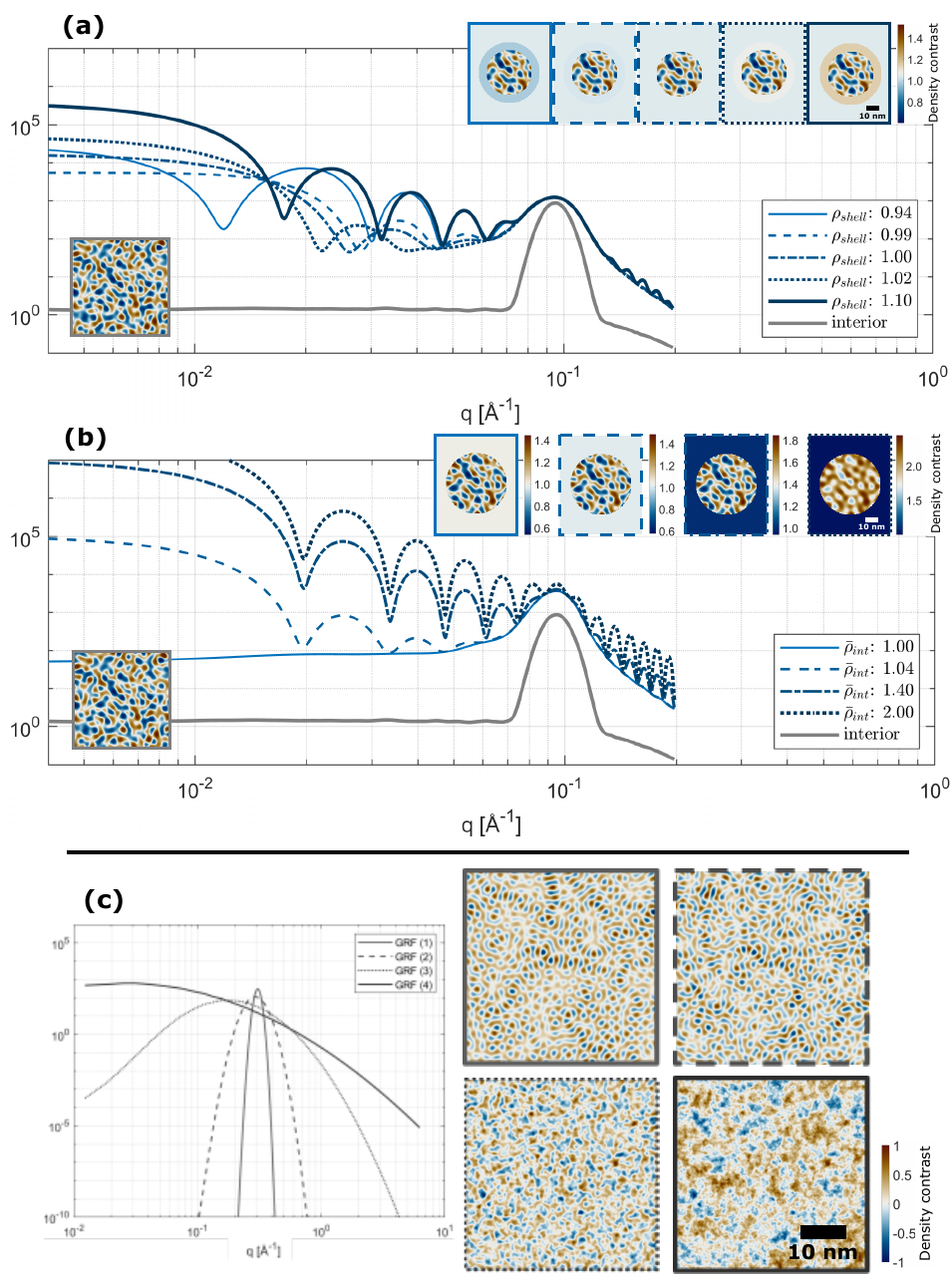}
    \caption{
    \textbf{SAXS curves for different electron-density configurations.}
    Illustration of how the simulated scattering profile responds to changes in (a) shell electron density \(\rho_{\mathrm{shell}}\), (b) the mean interior electron-density level \(\bar{\rho}_{\mathrm{int}}\), and (c) the GRF spectral parameters controlling interior disorder.
    Blue curves show SAXS curves computed from the full particle density \(\rho(\mathbf{x})\), while gray curves show curves computed from the corresponding interior fields \(\rho_{\mathrm{int}}(\mathbf{x})\) in \textbf{(a)}--\textbf{(b)} and from the underlying GRF \(G(\mathbf{x})\) in \textbf{(c)}.
    Each curve is averaged over 50 independent realizations.
    A framed central slice of the corresponding 3D electron density is shown for each configuration, with frame style matching the curve style.
    \textbf{(a)}--\textbf{(b)} The full particle model \(\rho(\mathbf{x})\) (blue) is shown together with the corresponding interior field \(\rho_{\mathrm{int}}(\mathbf{x})\) (gray) while varying \(\rho_{\mathrm{shell}}\) from 0.94 to 1.1 in \textbf{(a)} and \(\bar{\rho}_{\mathrm{int}}\) from 1 to 2 in \textbf{(b)}.
    \textbf{(c)} SAXS curves for four single-component GRF spectra with \(w=0\) (i.e., only \(P_1(q)\)).
    All four spectra share the location parameter \(q_{0,1}=0.0475\) and vary in width with \(s_{q,1} = 0.0075\), \(0.025\), \(0.125\) and \(0.5\), respectively.
    }
    \label{fig:SAXS_model_particle}
\end{figure}

Figure~\ref{fig:SAXS_model_particle} illustrates the simulation model and the resulting range of SAXS curves.
Panels~\textbf{(a)} and \textbf{(b)} show how the full model \(\rho(\mathbf{x})\) responds to changes in particle-level parameters while keeping the same interior field \(\rho_{\mathrm{int}}(\mathbf{x})\), highlighting the coupling between shell/geometry and interior structure.
Panel~\textbf{(c)} illustrates how varying the GRF spectrum changes the interior patterns generated by \(G(\mathbf{x})\) and, consequently, the associated scattering signatures, with increasing disorder as the width of the log-normal power spectrum increases.

In panels~\textbf{(a)}--\textbf{(b)}, the interior field \(\rho_{\mathrm{int}}(\mathbf{x})\) has a comparatively regular appearance due to the narrow GRF spectrum (gray curve peaked around \(q\approx 10^{-1}\)).
Accordingly, the full model curves (blue) are primarily dominated by the particle geometry at low \(q\), while the interior structure contributes most strongly around \(q\approx 10^{-1}\). Nevertheless, interactions between the particle and interior are coupled, and the interplay is visible across the full \(q\)-range.
Panel~\textbf{(a)} shows that the shell density \(\rho_{\mathrm{shell}}\) affects the SAXS curve at essentially all length scales except near \(q\approx 10^{-1}\).
Although the difference between the intermediate cases (\(\rho_{\mathrm{shell}}=0.99,\,1,\,1.02\)) is small compared with the variation inside the particle (\(\rho_{\mathrm{int}}(\mathbf{x}) \subset [-0.48,\,0.48]\)) and is therefore difficult to discern in the density slices, it produces clear and systematic differences in the SAXS curves.
Panel~\textbf{(b)} shows how the relative prominence of the feature around \(q\approx 10^{-1}\) changes with the mean interior density contrast, quantified here by \(\bar{\rho}_{\mathrm{int}}-1\).

The particle-level morphology also modifies the interior's scattering signature.
In panels~\textbf{(a)}--\textbf{(b)}, the peak around \(q\approx 10^{-1}\) is broader in the full-model curve (blue) than in the interior-field curve alone (gray).
A model that attributes this feature solely to the interior would therefore infer a broader distribution of interior length scales than is actually present.
This coupling between shell/geometry and interior structure means that interior-associated features in the total SAXS signal cannot be interpreted independently of the full model morphology, thereby breaking the scale-separation assumption commonly used in SAXS analysis.\looseness-1

\begin{table}[ht]
\centering
\caption{Key simulation and discretization settings used to generate monodisperse SAXS training curves. Here $N_f = N + 2p$.}
\label{tab:sim_settings}
\setlength{\tabcolsep}{5pt}
\small
\begin{tabular}{lcc}
\toprule
\textbf{Setting} & \textbf{Symbol} & \textbf{Value} \\
\midrule
3D GRF grid size (voxels per axis) & $N$ & 512 \\
Voxel size & $\Delta x$ & $5.0~\text{\AA}$ \\
2D FFT plane size after padding & $N_f$ & 1024 \\
2D zero-padding (per side) & $p$ & 256 \\
GRF realizations per parameter setting & $n$ & 50 \\
$q$-range & $[q_{\min},q_{\max}]$ & $[0.001,\,0.200]~\text{\AA}^{-1}$ \\
Number of $q$ samples & $N_q$ & 300 \\
$q$ sampling &  & uniform in $q$ \\
\bottomrule
\end{tabular}
\end{table}

\paragraph{Implementation and training dataset generation.}
The forward simulator and SAXS preprocessing pipeline were implemented in JAX to leverage \texttt{jit} compilation and efficient batching via \texttt{vmap}/\texttt{lax}.
Training data for the monodisperse surrogate were generated by sampling the bounded parameter domain (Table~\ref{tab:param_bounds}) using a Sobol low-discrepancy sequence.
We simulated $32{,}768$ Sobol samples for training and an additional $1{,}024$ Sobol samples for the held-out test set.
For each sampled parameter setting, the monodisperse intensity was estimated by averaging over $n=50$ independent GRF realizations (Results, section~\ref{sec:surrogate_performance}).
The resulting curves were exported and used to train the neural surrogate in PyTorch (section~\ref{sec:ml_models}).

\subsubsection{Neural surrogate of the monodisperse forward map}\label{sec:ml_models}

Each simulated training example corresponds to the \emph{monodisperse} (single-radius) scattering profile of an LNP with radius and structural parameters \((r,\theta)\).
We work in log-intensity space and define
\[
y_{\mathrm{mono}}^{(\mathrm{sim})}(q\mid r,\theta) \;=\; \log_{10} I_{\mathrm{mono}}^{(\mathrm{sim})}(q\mid r,\theta),
\]
evaluated on a fixed \(q\)-grid.
Direct regression from \((r,\theta)\) to \(y_{\mathrm{mono}}^{(\mathrm{sim})}(q\mid r,\theta)\) is challenging because small parameter perturbations can induce oscillatory \emph{phase} shifts in the scattering pattern rather than smooth amplitude changes.
To improve conditioning, we represent curves in a fixed orthonormal discrete cosine transform (DCT) basis and train a parameter-to-coefficient predictor, so that oscillatory phase shifts in $q$-space become smooth amplitude changes in frequency space.
  
\paragraph{Global data normalization.}
We use a tilde to denote global standardization of the dataset.
Inputs are min--max scaled using training-set bounds \(x_{\min},x_{\max}\).
Output log-intensity curves are standardized using fixed training-set statistics,
\begin{equation}
\tilde y_{\mathrm{mono}}^{(\mathrm{sim})}(q\mid r,\theta)
=
\frac{
y_{\mathrm{mono}}^{(\mathrm{sim})}(q\mid r,\theta)-y_{\mathrm{mean}}
}{
y_{\mathrm{std}}+\varepsilon
}.
\label{eq:global_standardization}
\end{equation}
The surrogate is trained to predict standardized curves \(\tilde y_{\mathrm{mono}}^{(\mathrm{sim})}\).
Predictions on the original log-intensity scale are obtained by inverting the standardization,
\begin{equation}
y_{\mathrm{mono}}^{(\mathrm{pred})}(q\mid r,\theta)
=
y_{\mathrm{std}}\ \tilde y_{\mathrm{mono}}^{(\mathrm{pred})}(q\mid r,\theta)
+
y_{\mathrm{mean}},
\label{eq:invert_global_standardization}
\end{equation}
and intensities on the original scale are recovered as
\(
I_{\mathrm{mono}}^{(\mathrm{pred})}(q\mid r,\theta)=10^{\,y_{\mathrm{mono}}^{(\mathrm{pred})}(q\mid r,\theta)}.
\)

\paragraph{Discrete cosine transform (DCT).}
Let \(\tilde y_j=\tilde y_{\mathrm{mono}}^{(\mathrm{sim})}(q_j\mid r,\theta)\) denote the standardized curve sampled at $N_q$ $q$ values \(q_j\) (\(j=0,\dots,N_q-1\)).
We use the orthonormal DCT-II basis matrix \(\Phi\in\mathbb{R}^{N_q\times N_q}\),
\begin{equation}
\Phi_{jk}
=
\alpha_k
\cos\!\left[
\frac{\pi}{N_q}\left(j+\tfrac{1}{2}\right)k
\right],
\qquad
\alpha_0 = \frac{1}{\sqrt{N_q}},\ \alpha_k = \sqrt{\frac{2}{N_q}}\ (k>0),
\label{eq:dct_basis}
\end{equation}
and retain only the AC components \(k\ge 1\), denoted \(\Phi_{\mathrm{AC}}\in\mathbb{R}^{N_q\times M}\).
Here \(M=N_q-1\), i.e., we remove only the DC component ($k=0$) and retain all $M = N_q - 1$ AC components. The DCT serves as a fixed orthonormal basis. Empirically, it reduces long-range output correlations and improves conditioning by representing oscillatory structures as smoothly varying coefficients rather than phase-shifted patterns in $q$-space.\looseness-1

\paragraph{Surrogate construction.}
We parameterize each standardized curve by (i) sample-specific affine parameters $\mu\in\mathbb{R}$ and $\sigma>0$, which capture the per-curve mean and scale, and (ii) an AC coefficient vector $h\in\mathbb{R}^{M}$, which captures the shape variation around that mean.
The \emph{parameter-to-DCT predictor} \(P_{\psi}\) is a residual neural network mapping
\[
P_{\psi}:(r,\theta)\ \mapsto\ (\mu,\log\sigma,h).
\]
The predicted curve in standardized space is reconstructed by the fixed DCT projection
\begin{equation}
\tilde y_{\mathrm{mono}}^{(\mathrm{pred})}(q\mid r,\theta)
=
\mu(r,\theta)
+
\sigma(r,\theta)\,\big[\Phi_{\mathrm{AC}}\, h(r,\theta)\big],
\qquad
\sigma=\exp(\log\sigma).
\label{eq:dct_recon}
\end{equation}
We emphasize the distinction between the \emph{global} dataset standardization in \eqref{eq:global_standardization} (fixed constants \(y_{\mathrm{mean}},y_{\mathrm{std}}\)) and the \emph{sample-specific} affine parameters \((\mu,\sigma)\) in \eqref{eq:dct_recon}, which are predicted by \(P_{\psi}\) and capture residual per-curve offset and amplitude in standardized space.

For clarity in the following, we denote the completely differentiable surrogate (returning log-intensity on the original scale) by
\begin{equation}
f_{\psi}(r,\theta)(q)\ :=\ y_{\mathrm{mono}}^{(\mathrm{pred})}(q\mid r,\theta)
\ =\ y_{\mathrm{std}}\ \tilde y_{\mathrm{mono}}^{(\mathrm{pred})}(q\mid r,\theta)+y_{\mathrm{mean}}.
\label{eq:full_surrogate}
\end{equation}

\paragraph{Training objective.}
The model is trained on simulated pairs \(\{(r_i,\theta_i),\tilde y_{\mathrm{mono}}^{(\mathrm{sim})}(q\mid r_i,\theta_i)\}\) using a Huber loss on the predicted log-intensity curves, normalized per-curve by the dynamic range of the target:
\begin{equation}
\mathcal{L}_{\mathrm{surr}}
= \mathrm{SmoothL1}\!\left(\frac{\hat{y}}{d_y},\,\frac{y}{d_y}\right),
\quad d_y = \max(y) - \min(y),
\label{eq:loss_surr}
\end{equation}
where $y$ is the simulated log-intensity target, $\hat{y}$ is the surrogate's prediction, and $d_y$ is the per-curve dynamic range. The per-curve normalization makes the loss insensitive to absolute curve amplitude, so curves at very different intensity scales contribute comparably during training.
The affine parameters $\mu$ and $\sigma$ are not explicitly supervised; they are learned implicitly through the reconstruction objective.
Full loss definition and hyperparameters are provided in Supplementary Section~S2.

\paragraph{Training details.}
The surrogate $P_\psi$ was trained in PyTorch on $32{,}768$ simulated monodisperse curves, with a held-out test set of $1{,}024$ curves generated from independent Sobol samples.
The predictor $P_\psi$ is a residual neural network with 9 residual blocks and a hidden width of 1024, predicting $(\mu,\log\sigma,h)$ as in \eqref{eq:dct_recon}.
Network parameters $\psi$ were learned with AdamW (batch size 1024) for $3.2\times 10^5$ parameter-update steps (equivalently 10{,}000 passes over the training set under this batching).

\subsubsection{Differentiable polydisperse fitting}\label{sec:parameter_fitting}

\paragraph{Polydisperse forward map.}
The pretrained surrogate $f_\psi(r,\theta)(q)$ predicts the \emph{monodisperse} log-intensity curve $y_{\mathrm{mono}}(q\mid r,\theta)=\log_{10} I_{\mathrm{mono}}(q\mid r,\theta)$ (Section~\ref{sec:ml_models}).
Because the surrogate was trained on monodisperse curves that had already been averaged over $n=50$ independent GRF realizations per parameter setting (Section~\ref{sec:saxs_simulations}), a single evaluation of $f_\psi$ returns a GRF-averaged curve, and the only integral that remains at fitting time is over the radius distribution.
To model size polydispersity, we define an ensemble-averaged intensity by numerical quadrature over a latent variable $u\in[0,1]$,
\begin{equation}
\hat I(q\mid \theta,\xi)
= \sum_{k=1}^{K} w_k\, 10^{\,f_{\psi}(r_k,\theta)(q)},
\qquad r_k = T_\xi(u_k),
\label{eq:ensemble_avg}
\end{equation}
where $\{(u_k,w_k)\}_{k=1}^K$ are fixed quadrature nodes and weights on $[0,1]$, and $T_\xi:[0,1]\to\mathbb{R}_+$ is a learnable, differentiable mapping parameterized by $\xi$ that induces an effective radius distribution.
We use $K=64$ quadrature nodes to jointly optimize the structural parameters $\theta$ and the size-distribution parameters $\xi=(m_r,s_r)$.
Because the nodes $u_k$ are fixed while the mapped radii $r_k=T_\xi(u_k)$ change with $\xi$, the integration is differentiable with respect to $\xi$.
Intensities, not log-intensities, are averaged with equal weights before any log-space loss is applied.

The synthetic polydisperse targets in Section~\ref{sec:synth_targets} are constructed without the surrogate.
Each target is generated by drawing $N=5000$ independent radii from the ground-truth size distribution and simulating a fresh GRF realization for every draw using the full forward model of Section~\ref{sec:saxs_simulations}.
The resulting Monte Carlo average must therefore cover both the radius distribution and the residual GRF variability, which is why the sample count is much larger than the $K=64$ used at fitting time.

\paragraph{Radius distribution (truncated log-normal).}
We model size polydispersity by a log-normal radius distribution truncated to
$r\in[r_{\min},r_{\max}]$ with $r_{\min}=100$~\AA\ and $r_{\max}=500$~\AA.
We optimize the physical-space mean and standard deviation $(m_r,s_r)$ (in \AA).
Internally, we use the log-space parameterization $\log r \sim \mathcal{N}(\mu_{\ln r},\sigma_{\ln r}^2)$,
obtained from $(m_r,s_r)$, and map fixed Gauss--Legendre nodes in CDF space to parameter-dependent physical radii via the inverse CDF: 
\[
\begin{aligned}
F_A &= F_{\rm LN}(r_{\min};\mu_{\ln r},\sigma_{\ln r}), \qquad
F_B = F_{\rm LN}(r_{\max};\mu_{\ln r},\sigma_{\ln r}),\\
u_k' &= F_A + u_k(F_B-F_A), \qquad
r_k=\exp\!\left(\mu_{\ln r}+\sigma_{\ln r}\Phi^{-1}(u_k')\right).
\end{aligned}
\]
where $F_{\rm LN}$ is the CDF of the untruncated log-normal, and $\Phi^{-1}$ is the standard normal quantile. Because the inverse CDF maps uniform nodes to the target distribution, all quadrature weights $w_k$ are equal.
Finally, radii are mapped to the surrogate input by $r_{01}=(r-r_{\min})/(r_{\max}-r_{\min})$.

\paragraph{Normalization.}
To reduce sensitivity to overall intensity scaling, we normalize each curve by the
$q^2$-weighted integrated intensity, also known as the scattering invariant \cite{stribeck2007slicetheorem},
\[
Q_w \propto \int_{q\in\mathcal{W}} q^2 I(q)\,dq,
\]
approximated on the discrete $q$ grid using the trapezoidal rule. We then define
\[
I_{\mathrm{PI}}(q) \;=\; \frac{I(q)}{Q_w}.
\]
This normalization is recomputed at every forward evaluation during fitting, so that the predicted and experimental curves are compared in shape rather than on an absolute scale. $\mathcal{W}$ is the same $q$-window used for fitting (after masking excluded points).

\paragraph{Optimization objective.}
We fit $(\theta,\xi)$ by minimizing a weighted sum of mean squared errors in log- and linear-intensity space,
\begin{equation}
\label{eq:loss_fit}
\mathcal{L}(\theta,\xi)
=
\lambda_{\log}\,\mathrm{MSE}\!\big(\log_{10} I_{\mathrm{PI}},\,\log_{10} I_{\mathrm{PI,exp}}\big)
+
\lambda_{\mathrm{lin}}\,\mathrm{MSE}\!\big(I_{\mathrm{PI}},\,I_{\mathrm{PI,exp}}\big),
\end{equation}
where all terms are evaluated on the fixed $q$ grid with uniform weighting over $q$. The log-space term emphasizes weak high-$q$ features and relative deviations across the dynamic range, whereas the linear-space term prevents the fit from being dominated by small-intensity regions and stabilizes agreement at low $q$ with $\lambda_{\log} = 0.1$ and $\lambda_{\mathrm{lin}} = 1$.

\paragraph{Multi-start optimization, constraints, and retention.}
Because the objective in \eqref{eq:loss_fit} is non-convex, we use multi-start gradient-based optimization to reduce sensitivity to initialization and identify alternative near-optimal solutions.
Initial parameters \((\theta,\xi)\) are drawn from a scrambled Sobol sequence within the bounds of Table~\ref{tab:param_bounds}, sampled in min--max space and mapped to physical units.
To avoid starting at hard bounds, Sobol points are sampled from intervals that exclude the outer 5\% of each parameter range.
For GRF spectral locations, we enforce an ordering constraint by reparameterizing \(m_{q,2}\) with an auxiliary \(\eta\in(0,1)\) (see Supplementary section~S3), ensuring \(m_{q,2}\ge m_{q,1}+\Delta_{\min}\) throughout optimization without projections or penalties.
We run \(N_{\mathrm{runs}}=1024\) restarts per synthetic target and 8000 for the experimental SAXS profile. All other settings remain the same.
Each restart minimizes \eqref{eq:loss_fit} using AdamW with a cosine learning-rate schedule and a fixed step budget.
For efficiency, restarts are evaluated in parallel using a batched implementation that preserves the objective and retention criteria.
Although optimization uses the composite loss, we score and retain solutions based on the final \(\mathrm{RMSE}_{\log}\) between \(\log_{10} I_{\mathrm{PI}}(q)\) and \(\log_{10} I_{\mathrm{PI,exp}}(q)\) on the fixed \(q\) grid.
The retention criterion combines a robust threshold on \(\mathrm{RMSE}_{\log}\) with ensemble size bounds, as detailed in section~\ref{sec:eval_metrics}.
The retained ensemble is used for downstream correlation, clustering, and sensitivity analyses.
We interpret this ensemble as an empirical characterization of non-uniqueness in the inverse problem under the chosen forward model, not as a probabilistic posterior.\looseness-1

\paragraph{Analytical baseline models.}
To contextualize the GRF results, the same differentiable fitting framework was applied to two analytical baselines: a polydisperse homogeneous sphere and a polydisperse core--shell sphere.
The sphere form factor is $I(q,r) = r^6 f(qr)^2$ with $f(x) = 3(\sin x - x\cos x)/x^3$.
The core--shell amplitude is $F(q) = R^3 f(qR) + r_c^3\,(c-1)\,f(qr_c)$, where $R$ is the outer radius, $r_c = R - d$ is the core radius, $d$ is the shell thickness, and $c = \Delta\rho_{\rm core}/\Delta\rho_{\rm shell}$ is the contrast ratio. The overall scale cancels under Porod-invariant normalization.
Both baselines use the same truncated log-normal radius distribution and 64-point Gauss--Legendre quadrature as the GRF model.
Each was fitted at three $q$ cutoffs ($q_{\max} \in \{0.02,\,0.05,\,0.10\}$~\AA$^{-1}$) with 4096 Sobol-initialized restarts and the same composite loss in \eqref{eq:loss_fit}, so that differences in inferred parameters reflect the forward model rather than the fitting procedure.
Full curves over the complete $q$ range are evaluated at every restart regardless of the fitting cutoff.
Results and detailed diagnostics are reported in Supplementary section~S8.

\begin{table}[tbp]
\centering
\caption{Parameter bounds used during optimization.}
\label{tab:param_bounds}

\small
\setlength{\tabcolsep}{5pt}
\renewcommand{\arraystretch}{1.12}

\begin{tabular}{@{}lccp{10.0cm}@{}}
\toprule
\textbf{Quantity} & \textbf{Min} & \textbf{Max} & \textbf{Description}\\
\midrule
$[r_{\min},r_{\max}]$ (Å) & 100 & 500 & Fixed support for the radius distribution and surrogate normalization \\
$m_r$ (Å) & 120 & 350 & Mean radius (physical space) \\
$s_r$ (Å) & 10 & 100 & Radius standard deviation (physical space) \\
$\Delta\rho_{\mathrm{int}}$ & -1.0 & 1.0 & GRF density offset \\
$m_{q,1}$ & 0.05 & 0.15 & GRF spectrum location (component 1) \\
$m_{q,2}$ & 0.05 & 0.15 & GRF spectrum location (component 2) \\
$w$ & 0.0 & 1.0 & Mixture weight between spectral components \\
$s_{q,1}$ & 0.01 & 0.15 & GRF spectrum width (component 1) \\
$s_{q,2}$ & 0.01 & 0.15 & GRF spectrum width (component 2) \\
$d$ (Å) & 40 & 70 & Shell thickness \\
$\rho_{\mathrm{rel}}$ & 0.7 & 1.3 & Relative shell density \\
\bottomrule
\end{tabular}
\end{table}

\subsubsection{Synthetic polydisperse benchmark construction}\label{sec:synth_targets}

To evaluate inference under controlled conditions, we construct synthetic polydisperse SAXS targets with known ground-truth parameters.
For each target, we parameterize the radius distribution as a truncated log-normal and keep the structural parameters $\theta$ fixed within the target while varying the distribution parameters across targets.

We draw radii $r^{(s)} \sim p(r\mid\xi)$ and simulate an independent GRF-based particle realization for each draw using the forward model in section~\ref{sec:saxs_simulations}.
The target curve is obtained by Monte Carlo averaging over $N=5000$ particles,
\[
I_{\mathrm{target}}(q)=\frac{1}{N}\sum_{s=1}^{N} I_{\mathrm{mono}}\!\left(q \mid r^{(s)},\theta\right).
\]
With $N=5000$, residual Monte Carlo variability is negligible for the analyses reported here, so the synthetic targets can be treated as effectively noise-free.
The ground-truth parameters for all four test cases are listed in Supplementary Table~S2.

\subsubsection{Analysis of retained solutions}\label{sec:solution_analysis}
We analyze the structural identifiability of the parameterization by examining correlations, clustering, and local sensitivities within the near-optimal set.

\paragraph{Correlation analysis.}
To quantify dependencies within the retained near-optimal ensemble $\mathcal{K}$, we compute Spearman rank correlations for all parameter pairs. Unless stated otherwise, correlations are computed on the same parameter representation used during optimization (min--max scaled to $[0,1]$). To emphasize the best-fitting runs within $\mathcal{K}$, we use loss-based weights
\[
w_i \propto \exp(-L_i/s), \qquad s=\mathrm{std}(\{L_i\}_{i\in\mathcal{K}}),
\]
where $L_i$ is the final $\mathrm{RMSE}_{\log}$ for restart $i$. We visualize the ten pairs with the largest absolute weighted correlations using hexbin density plots.

\paragraph{Clustering and representative solutions.}
To identify distinct modes within the retained ensemble, we cluster solutions in parameter space.
We first standardize the retained parameter vectors (zero mean, unit variance computed over $\mathcal{K}$) using a \texttt{StandardScaler}.
We then apply HDBSCAN (Euclidean metric) with \texttt{min\_cluster\_size}=100 and \texttt{min\_samples}=30.
HDBSCAN identifies dense clusters without requiring the number of clusters to be specified and assigns non-members to a noise class.
For visualization, we compute a two-dimensional PCA embedding of the standardized vectors and color points by cluster label (Fig.~\ref{fig:clustering}a). Cluster-resolved marginals are computed by restricting to members of each cluster (Fig.~\ref{fig:clustering}b).

For each cluster, we select a single representative solution as the medoid in the standardized space (the solution with the minimum average Euclidean distance to other cluster members). Ties are broken by lower loss.
For each representative parameter set, we generate a corresponding real-space electron-density field by sampling one GRF realization at the fitted structural parameters and visualizing a central 2D slice.
The associated radius distribution is obtained directly from the fitted polydispersity parameters (truncated log-normal, Fig.~\ref{fig:poly}).

\paragraph{Sensitivity and local identifiability.}
For each retained solution, we compute the Jacobian of the predicted $q^2$-normalized log-intensity curve with respect to the fitted parameters. Parameters are min--max scaled to $[0,1]$ using the same bounds as in optimization to obtain dimensionless sensitivities. Let $\tilde{\theta}$ denote the scaled parameters. We define
\[
J_{ij}=\frac{\partial \log_{10} I(q_i;\theta)}{\partial \tilde{\theta}_j}.
\]
To summarize sensitivities across the retained ensemble, we aggregate Jacobians by the elementwise median of $J$ (signed). Global sensitivity scores are computed as the RMS of each parameter's aggregated Jacobian profile over $q$. To characterize local coupling, we compute the cosine similarity between aggregated Jacobian profiles for parameter pairs $a,b$,
\[
C_{ab}=\frac{\langle J_{:a},J_{:b}\rangle}{\|J_{:a}\|\;\|J_{:b}\|}.
\]
As an additional curvature-based summary, we report the Gauss--Newton matrix $F=J^\top J$ evaluated at each retained solution.

\subsubsection{Evaluation metrics}\label{sec:eval_metrics}

We assess surrogate accuracy and fit quality using metrics computed in $\log_{10}$-intensity space on the fixed $q$ grid.
For a predicted curve $I_{\mathrm{pred}}(q)$ and reference $I_{\mathrm{ref}}(q)$, we define the signed log-residual
\begin{equation}
\Delta_{\log}(q)=\log_{10} I_{\mathrm{pred}}(q)-\log_{10} I_{\mathrm{ref}}(q).
\end{equation}
Surrogate accuracy is summarized by the mean absolute logarithmic error (MALE),
\begin{equation}
\mathrm{MALE}_{\log}=\frac{1}{N_q}\sum_{i=1}^{N_q}\left|\Delta_{\log}(q_i)\right|,
\end{equation}
and a dynamic-range-normalized relative deviation,
\begin{equation}
\mathrm{RelErr}_{\log}=
\frac{\sum_i |\Delta_{\log}(q_i)|}{\max_q \log_{10} I_{\mathrm{ref}}(q)-\min_q \log_{10} I_{\mathrm{ref}}(q)}.
\end{equation}
For thresholding near-optimal fits in multi-start optimization, we use the root-mean-square error in log space,\looseness-1
\begin{equation}
\mathrm{RMSE}_{\log}=
\sqrt{\frac{1}{N_q}\sum_{i=1}^{N_q}\Delta_{\log}(q_i)^2},
\end{equation}
and denote the final value for restart $i$ by $L_i$.

\paragraph{Selection of near-optimal solutions.}
Although optimization minimizes the composite loss in \eqref{eq:loss_fit}, we define the retained \emph{near-optimal} set using the interpretable curve-space metric $\mathrm{RMSE}_{\log}$.
Let $L_i$ denote the final $\mathrm{RMSE}_{\log}$ for restart $i=1,\dots,N$.
To avoid sensitivity to a single best restart, we use a robust reference loss $L_\star$ defined as the mean of the lowest $\lceil p_{\mathrm{best}}\,N\rceil$ values of $L_i$ (with $p_{\mathrm{best}}=0.10$).
We first form the acceptance set
\begin{equation}
\mathcal{A}_\alpha=\{\, i : L_i \le \alpha\,L_\star\,\},
\end{equation}
with $\alpha=1.2$ unless stated otherwise.
To avoid retaining too few or too many solutions, we enforce ensemble-size bounds
\begin{equation}
K_{\min}=\lceil \beta_{\min}N\rceil,\qquad K_{\max}=\lceil \beta_{\max}N\rceil,
\end{equation}
with $\beta_{\min}=0.05$ and $\beta_{\max}=0.30$.
If $|\mathcal{A}_\alpha|<K_{\min}$, we retain the $K_{\min}$ lowest-loss restarts. If $|\mathcal{A}_\alpha|>K_{\max}$, we retain the $K_{\max}$ lowest-loss restarts. Otherwise, we retain $\mathcal{A}_\alpha$.
The resulting index set is denoted $\mathcal{K}$.

\section{Data availability}
\label{sec:data}
The data supporting the findings of this study are openly available at Zenodo, \url{https://doi.org/10.5281/zenodo.20338599}. The deposit contains the experimental MC3 LNP SAXS curve, the four synthetic benchmark targets, the simulated data used to train the neural surrogate, and the pretrained surrogate weights.

\section{Code availability}
\label{sec:code}
The source code and trained model weights are available at \url{https://github.com/mariabankestad/aisaxs} and archived on Zenodo at \url{https://doi.org/10.5281/zenodo.20338599}.

\section{Acknowledgements}
\label{sec:acknowledgements}
This work was funded by Vinnova through the project AI-SAXS: Decoding Structural Complexity with Intelligent Scattering Analysis, grant no.~2023-02701.
We thank MAX IV Laboratory and the CoSAXS beamline staff, in particular Marc Obiols, for support during the SAXS measurements.
This work benefited from the use of the SasView application, originally developed under NSF award DMR-0520547. SasView contains code developed with funding from the European Union's Horizon 2020 research and innovation program under the SINE2020 project, grant agreement No.~654000.
The computations were enabled by resources provided by the National Academic Infrastructure for Supercomputing in Sweden (NAISS), partially funded by the Swedish Research Council through grant agreement no.~2022-06725.

\section{Author contributions}
\label{sec:contrib}
%
M.B.: Conceptualization, Methodology, Software, Formal analysis, Validation, Visualization, Writing -- original draft, Writing -- review \& editing.
S.B.: Conceptualization, Methodology, Software, Formal analysis, Validation, Writing -- review \& editing.
M.R.: Conceptualization, Methodology, Software, Formal analysis, Funding acquisition, Writing -- review \& editing
E.K.: Conceptualization, Methodology, Funding acquisition, Writing -- review \& editing
V.M.: Investigation, Data curation, Resources, Writing -- review \& editing
A.G.: Investigation, Data curation, Resources, Writing -- review \& editing
M.M.: Investigation, Data curation, Resources, Writing -- review \& editing
M.Y.A.: Investigation, Data curation, Resources, Writing -- review \& editing
S.N.: Investigation, Data curation, Resources, Funding acquisition, Writing -- review \& editing
A.T.: Investigation, Data curation, Funding acquisition, Writing -- review \& editing
S.C.: Writing -- review \& editing
S.Y.: Funding acquisition, Software, Formal analysis, Investigation, Writing -- review \& editing
J.R.: Funding acquisition, Writing -- review \& editing
S.P.: Funding acquisition, Writing -- review \& editing

\section{Competing interests}
\label{sec:conflict}
M.R., E.K., V.M., A.G., M.M., M.Y.A., and S.N. are employees of AstraZeneca and have stock ownership and/or stock options or interests in the company. All other authors declare no financial or non-financial competing interests.

\bibliography{refs}

\clearpage
\appendix
\renewcommand{\thesection}{S\arabic{section}}
\renewcommand{\thesubsection}{\thesection.\arabic{subsection}}
\renewcommand{\thesubsubsection}{\thesubsection.\arabic{subsubsection}}
\setcounter{section}{0}
\renewcommand{\thefigure}{S\arabic{figure}}
\setcounter{figure}{0}
\renewcommand{\thetable}{S\arabic{table}}
\setcounter{table}{0}
\renewcommand{\theequation}{S\arabic{equation}}
\setcounter{equation}{0}

\begin{center}
{\Large\bfseries Supplementary Information}\\[3pt]
\textit{A differentiable machine learning small-angle X-ray scattering analysis framework for structure elucidation of lipid nanoparticles}
\end{center}

\vspace{1em}


\section{Samples from the training data}

\begin{figure}[htbp]
  \centering
  \includegraphics[width=0.8\linewidth]{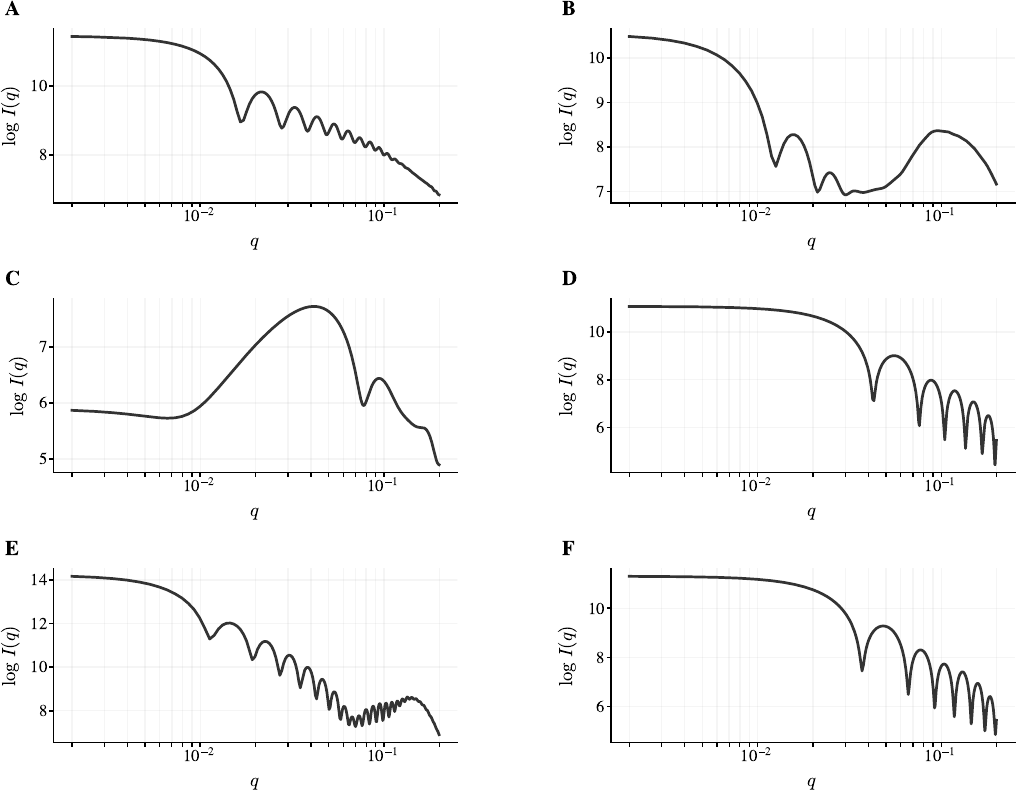}
  \caption{%
    \textbf{Representative training curves.}
    Six SAXS scattering curves from the training set, selected by greedy farthest-point
    sampling to illustrate the range of curve shapes present in the data.
    Each panel shows $\log I(q)$ versus $q$ on a logarithmic $q$-axis.
    Panels are labelled \textbf{A}--\textbf{F} in order of decreasing shape-space distance
    from the initial seed curve.
  }
  \label{fig:diversity}
\end{figure}

Figure~\ref{fig:diversity} illustrates the diversity of SAXS curve shapes in the training data.
Six representative curves were selected by greedy farthest-point sampling in shape space: starting from a random curve, each successive pick is the training sample whose mean-subtracted log-intensity profile is maximally distant from all already-selected curves.
The selection covers a broad range of profiles, including pronounced low-$q$ plateaus, steep power-law decays, and oscillatory features at intermediate $q$.
This diversity reflects the range of structural configurations spanned by the parameter domain and motivates the need for an output representation that can reproduce a variety of curve morphologies while producing physically smooth predictions.

\section{Surrogate training: loss function and hyperparameters}
\label{sec:s1_loss}

The surrogate is trained on simulated pairs $\{(r_i,\theta_i),\, \tilde{y}_{\mathrm{mono}}^{(\mathrm{sim})}\}$ using a Huber (SmoothL1) loss on the predicted log-intensity curves, normalized per-curve by the dynamic range of the target:
\begin{equation}
\mathcal{L}_{\mathrm{surr}}
= \mathrm{SmoothL1}\!\left(\frac{\hat{y}}{d_y},\,\frac{y}{d_y}\right),
\qquad d_y = \max(y) - \min(y),
\label{eq:loss_surr_full}
\end{equation}
where $y$ is the log-intensity target, $\hat{y}$ is the surrogate's prediction, and $d_y$ is the per-curve dynamic range over the $N_q$ grid points. The per-curve normalization is the central design choice: it makes the loss insensitive to absolute curve amplitude, so that curves with very different intensity scales contribute comparably during training and the gradient is not dominated by curves with the largest dynamic range. The Huber form (PyTorch \texttt{SmoothL1Loss} with default $\delta$) is quadratic for small residuals and linear for large ones, providing robustness to occasional outlying training targets without discarding them entirely. The remaining training hyperparameters (architecture, optimizer settings, batch size, and hardware) are listed in Table~\ref{tab:hyperparams}.

\begin{table}[ht]
\centering
\caption{Surrogate training hyperparameters.}
\label{tab:hyperparams}
\begin{tabular}{lll}
\toprule
Hyperparameter & Value & Notes \\
\midrule
Architecture         & 9 residual blocks, 1024 units each & SiLU activations \\
Input dimension      & $d_\theta = 9$                 & \\
Output dimension     & $d_z = 301$                      & $\mu$, $\log\sigma$, $h$ (DCT-AC) \\
Loss                 & Huber (SmoothL1)                & Log-intensity space \\
Optimizer            & AdamW                           & lr cosine $10^{-3} \to 10^{-7}$ \\
Batch size           & 1024                            & With gradient accumulation \\
Epochs               & 10000                           & \\
Hardware             & NVIDIA A100 (80\,GB)            & Training time $\approx$ 2\,h \\
\bottomrule
\end{tabular}
\end{table}

\section{Ordering constraint reparameterization}
\label{sec:s2_repar}

The GRF power spectrum is parameterized as a mixture of two log-normal components centered at $m_{q,1}$ and $m_{q,2}$ (main text, electron-density synthesis section).
Without an ordering constraint, gradient-based optimization can drive $m_{q,2} \le m_{q,1}$, making the two components indistinguishable and breaking the intended two-scale interpretation.

Rather than enforcing $m_{q,2} > m_{q,1}$ with a projection or a penalty term, we reparameterize $m_{q,2}$ via an unconstrained scalar $\eta_{\mathrm{raw}} \in \mathbb{R}$.
Let $h = 0.15$~\AA$^{-1}$ be the upper bound on both $m_{q,1}$ and $m_{q,2}$, and let $\varepsilon = 10^{-4}$ be a small numerical safety margin.
The effective fraction $\eta$ is obtained by a squashed sigmoid:
\begin{equation}
\eta = \varepsilon + (1 - 2\varepsilon)\,\sigma(\eta_{\mathrm{raw}}),
\qquad \sigma(x) = \frac{1}{1+e^{-x}},
\label{eq:eta_map}
\end{equation}
so that $\eta \in (\varepsilon,\,1-\varepsilon)$ for all finite $\eta_{\mathrm{raw}}$.
The constrained spectral location is then
\begin{equation}
m_{q,2} = m_{q,1} + \eta\,(h - m_{q,1}).
\label{eq:mq2_repar}
\end{equation}
Because $\eta > 0$ and $h > m_{q,1}$ by construction, Eq.~\eqref{eq:mq2_repar} guarantees $m_{q,2} > m_{q,1}$ throughout optimization without any projection step or penalty.
Gradient flow through both equations is smooth: $\partial m_{q,2}/\partial \eta_{\mathrm{raw}}$ is nonzero everywhere and passes through the sigmoid and the linear combination.

During multi-start optimization, each restart maintains its own $\eta_{\mathrm{raw}}$, initialized from the Sobol sequence by inverting Eqs.~\eqref{eq:eta_map}--\eqref{eq:mq2_repar}.

\section{Ablation study: neural network output architectures}
\label{sec:s3_ablation}

Predicting a full SAXS intensity profile from a small set of physical parameters is a challenging regression problem: small parameter perturbations can induce oscillatory phase shifts in the scattering pattern rather than smooth amplitude changes.
We compare two neural network architectures for this task: a \emph{Direct MLP} that outputs the full scattering curve directly, and a \emph{DCT model} that outputs coefficients in a discrete cosine transform basis before reconstructing the curve.
The DCT model enforces smoothness in the output by representing the curve through a fixed orthonormal basis, while the Direct MLP places no structural constraint on its predictions.

The comparison uses the same $32{,}768$ simulated SAXS curves used to train the production surrogate (Section~\ref{sec:s1_loss}).
Six representative curves from this set, selected by greedy farthest-point sampling in mean-subtracted log-intensity space, are shown in Figure~\ref{fig:diversity}.

\begin{figure}[htbp]
  \centering
  \includegraphics[width=0.7\linewidth]{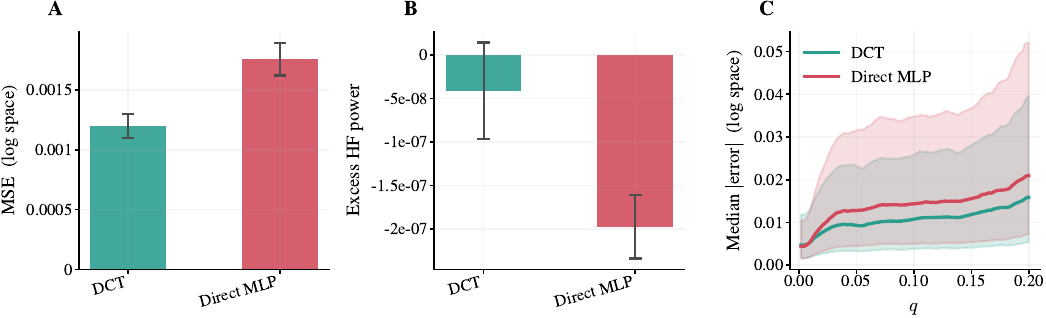}
  \caption{%
    \textbf{Quantitative comparison of DCT and Direct MLP on the external test set.}
    Results are averaged over five random train/validation splits; error bars show the
    standard deviation across splits.
    \textbf{(A)}~Mean squared error in log-intensity space.
    \textbf{(B)}~Excess high-frequency power.
    \textbf{(C)}~Median absolute error as a function of $q$ (shaded: 10--90\% band).
  }
  \label{fig:metrics}
\end{figure}

Figure~\ref{fig:metrics} shows quantitative results averaged over five random 90/10 train/validation splits of the $32{,}768$ samples.
Panel~A shows the mean squared error (MSE) in log-intensity space.
Panel~B shows an \emph{excess high-frequency power} metric: the fraction of the true curve's total spectral power by which the predicted curve exceeds the true curve in the upper 50\% of DCT frequencies.
A positive value indicates spurious oscillations not present in the ground truth.
Panel~C shows the median absolute error as a function of $q$, with the shaded band spanning the 10th to 90th percentile.

\begin{figure}[htbp]
  \centering
  \includegraphics[width=0.7\linewidth]{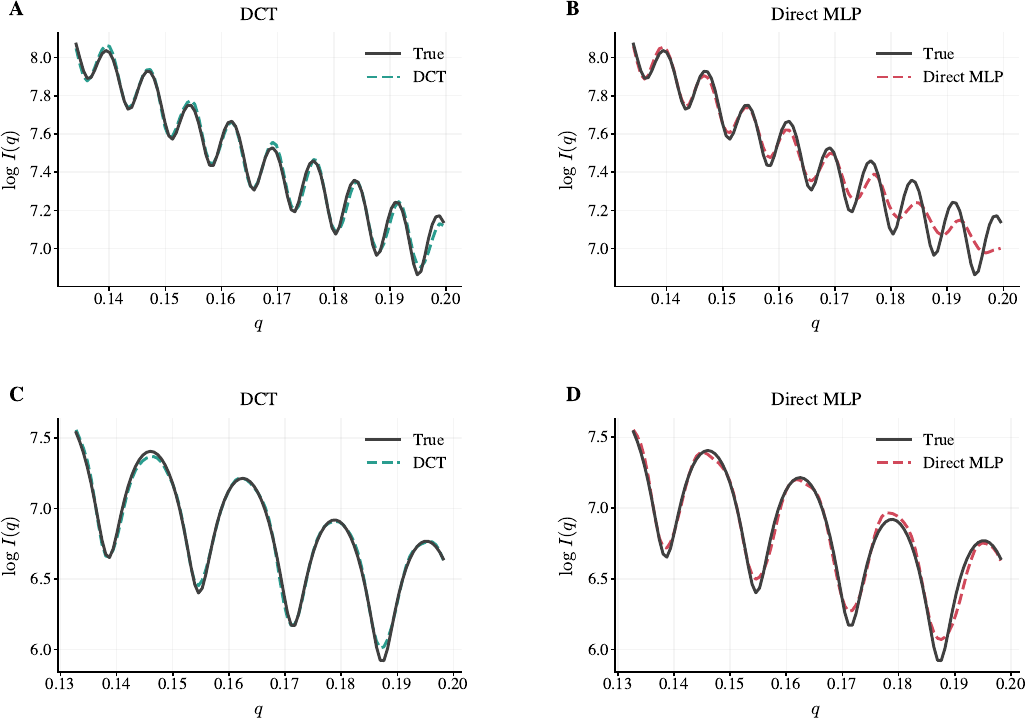}
  \caption{%
    \textbf{Prediction examples illustrating the smoothness difference.}
    Two example curves (rows), each shown for the DCT model (left, \textbf{A},~\textbf{C})
    and the Direct MLP (right, \textbf{B},~\textbf{D}).
    The true curve is shown in dark gray; the model prediction is a dashed colored line.
  }
  \label{fig:wiggle}
\end{figure}

Both models achieve similar MSE (Figure~\ref{fig:metrics}A), but the Direct MLP produces physically implausible high-frequency oscillations in a subset of predictions.
Figure~\ref{fig:wiggle} shows two example predictions for each architecture.
The DCT model produces smooth, physically plausible curves in both cases, whereas the Direct MLP introduces spurious oscillations absent from the ground truth.
These oscillations would propagate into the polydisperse intensity computation and the gradient-based fitting, potentially introducing artifacts in the inferred parameters.
The DCT architecture is therefore used for all results in the main text.

\section{Surrogate robustness across the parameter domain}
\label{sec:s4_robustness}

The identifiability analyses in the main text require the surrogate to be accurate not just on average but across the full parameter domain, since multi-start optimization explores diverse regions of the parameter space.
If the surrogate were systematically less accurate in some region, near-optimal solutions in that region could reflect surrogate error rather than genuine forward-model degeneracy.

To assess this, we stratify the held-out test set (1024 samples) by each parameter individually, binning samples into quantiles by the parameter value.
Within each bin, we compute the median curve-space error ($\mathrm{RMSE}_{\log}$) together with the interquartile range.
All other parameters vary according to the joint test distribution in each bin, so the error profiles reflect the surrogate's accuracy conditional on one parameter while marginalizing over the rest.

\begin{figure}[h!]
    \centering
    \includegraphics[width=0.8\linewidth]{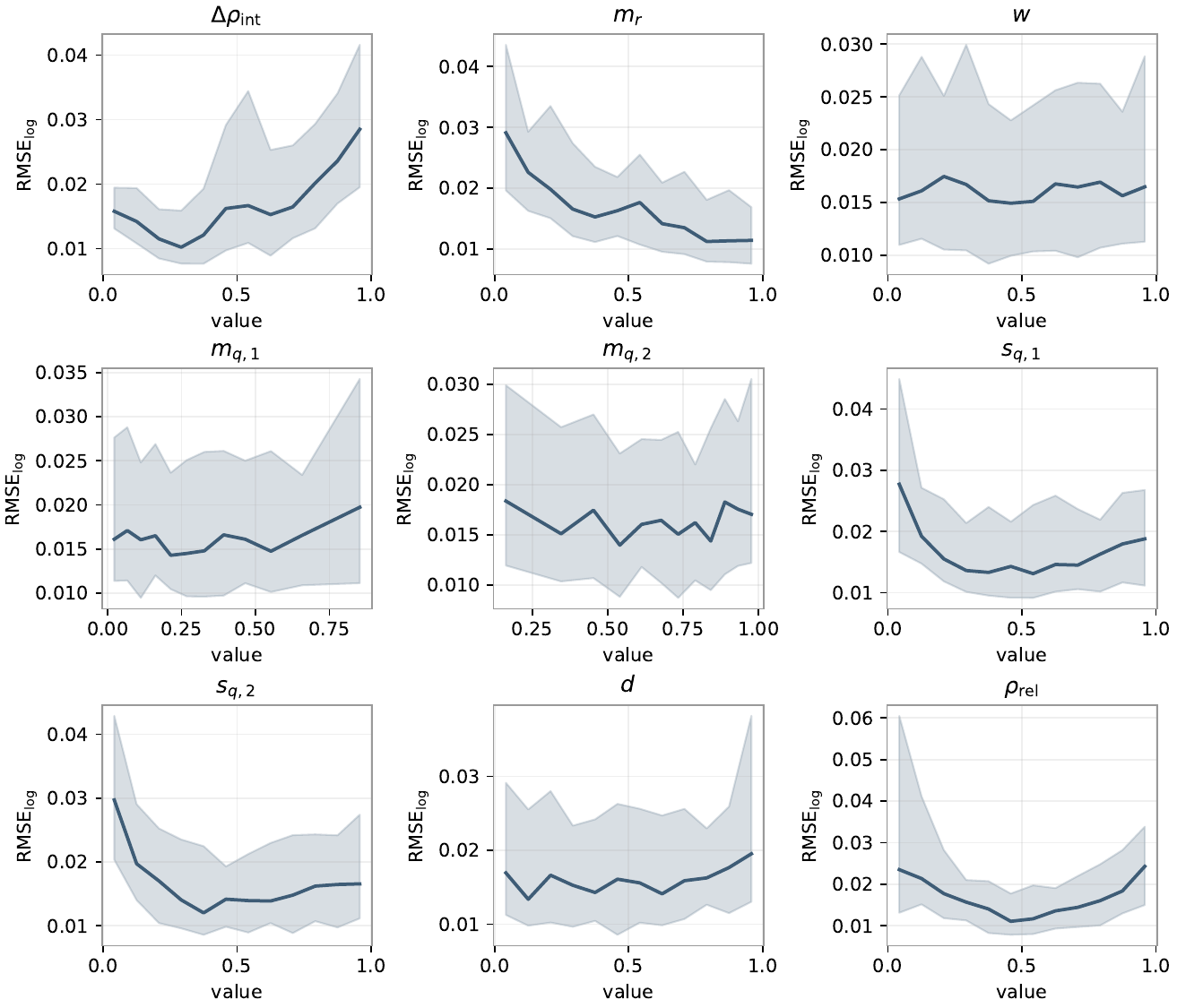}
    \caption{
    Robustness of the surrogate across the parameter domain, evaluated on a held-out test set of 1024 samples.
    For each parameter, the test-set samples are stratified into quantile bins by the parameter value.
    Within each bin, we report the median curve-space error ($\mathrm{RMSE}_{\log}$) together with the interquartile range (shaded band).
    Flat error profiles indicate uniform generalization across the sampled range, whereas systematic increases toward the domain boundaries reveal regions where predictive accuracy degrades.
    All other parameters vary according to the joint test distribution in each panel.
    }
    \label{fig:parameter_robustness}
\end{figure}

Figure~\ref{fig:parameter_robustness} shows the results.
The error profiles are approximately flat for most parameters, indicating that the surrogate generalizes uniformly across their sampled ranges.

\section{Quadrature convergence for polydispersity integration}
\label{sec:s5_quadrature}

The polydisperse SAXS intensity is computed by integrating the monodisperse form factor over the log-normal radius distribution using $K$-point Gauss--Legendre quadrature.
Because the integrand is the composition of a smooth form factor with the inverse CDF of a truncated log-normal, it is analytic and rapidly decaying, and Gauss--Legendre quadrature achieves exponential convergence in $K$ for this class of integrands.

To confirm that $K = 64$ (as used throughout the paper) is well within the converged regime, we ran a systematic study across 200 random parameter configurations.
Each configuration was drawn from the production prior using a Sobol sequence: GRF spectral parameters were sampled uniformly in $[0,1]$ (model-normalized space), while the mean radius was drawn from $m_r \in [150, 270]\,\text{\AA}$ and the coefficient of variation from $\mathrm{cv} \in [0.10, 0.35]$, matching the bounds used during experimental fitting.
For each configuration, we evaluated the polydisperse curve at $K \in \{4, 8, 16, 32, 64, 128\}$ and at a high-accuracy reference $K_{\mathrm{ref}} = 256$, then computed the root-mean-square error in $\log_{10}\,I(q)$ over the full $q$ range:
\begin{equation}
    \mathrm{RMSE}_{\log}(K) \;=\;
    \sqrt{\frac{1}{N_q} \sum_{j=1}^{N_q}
        \bigl(\log_{10} I_K(q_j) - \log_{10} I_{256}(q_j)\bigr)^2}.
\end{equation}

\begin{figure}[ht]
    \centering
    \includegraphics[width=0.7\linewidth]{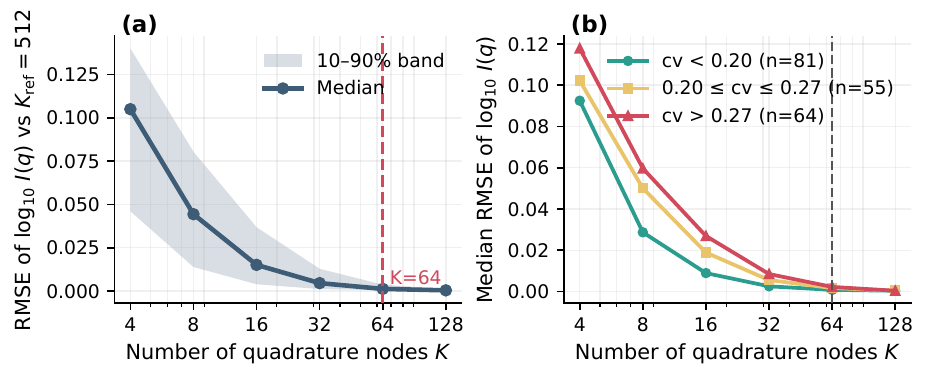}
    \caption{
        \textbf{Quadrature convergence for polydispersity integration.}
        \textbf{(a)}~$\mathrm{RMSE}_{\log}$ of the polydisperse SAXS curve
        (versus a $K_{\mathrm{ref}} = 256$ reference) as a function of the
        number of Gauss--Legendre nodes $K$, across 200 random parameter
        configurations.
        Shaded band: 10th--90th percentile; solid line: median.
        The vertical dashed line marks $K = 32$.
        \textbf{(b)}~Median $\mathrm{RMSE}_{\log}$ stratified by coefficient of
        variation ($\mathrm{cv}$), showing that convergence at $K = 32$ is
        consistent across the full polydispersity range used in the paper.
    }
    \label{fig:quadrature_convergence}
\end{figure}

Figure~\ref{fig:quadrature_convergence} shows the results.
Panel~(a) plots the median and 10th--90th percentile band of $\mathrm{RMSE}_{\log}$ across all 200 configurations as a function of $K$.
The error falls steeply between $K = 4$ and $K = 16$.
By $K = 32$ (dashed line in figure), the median error is already small, and by $K = 64$ it is below $\num{0.002}$ in $\log_{10}$ units.
$K = 128$ produces negligible further improvement.
Panel~(b) stratifies the same curves by polydispersity group ($\mathrm{cv} < 0.20$, $0.20 \leq \mathrm{cv} \leq 0.27$, $\mathrm{cv} > 0.27$), confirming that convergence holds uniformly across the full range of polydispersity values used in the paper.
The paper uses $K = 64$, which provides a comfortable margin beyond the onset of convergence at $K = 32$.

\section{Synthetic benchmark construction}
\label{sec:s7_synthetic}

The main text describes the Monte Carlo averaging procedure used to generate each synthetic target (Methods, ``Synthetic polydisperse benchmark construction'').
This section provides the full ground-truth parameter table and implementation details needed to reproduce the benchmark.

Four noiseless polydisperse SAXS targets were constructed from the physical LNP model.
Each target curve is the mean over $N=5000$ independent particle simulations, one fresh GRF realization per particle, on a uniform $q$-grid of 300 points spanning $q \in [0.001,\,0.2]$\,\AA$^{-1}$.
At $N=5000$, residual Monte Carlo variability is negligible, and the targets can be treated as noise-free ground truth.

When the inference pipeline is applied to a synthetic target, all parameters are optimized jointly: the structural GRF parameters $\theta$ (spectral mode weights $w_q$, positions $m_q$, widths $s_q$, shell thickness, shell contrast, GRF shift) and the size distribution parameters $\xi$ (mean radius $m_r$, standard deviation $s_r$), with radii constrained to $[100,\,500]\,\text{\AA} $.
During the surrogate forward pass, a single GRF realization is drawn for each radius sample, matching the computational budget used for the experimental data.

Table~\ref{tab:synth_cases} lists the ground-truth parameters for all four test cases.
The cases span mean radii from 190 to 260\,\AA, polydispersity standard deviations from 25 to 65\,\AA, and a range of interior structural contrasts, providing a varied benchmark for assessing inference accuracy.

\begin{table}[ht]
\centering
\caption{%
  Ground-truth parameters for the four synthetic benchmark targets.
  $w_q$, $m_q$, $s_q$ are the weight, position, and width of each spectral mode of the
  GRF power spectrum; $m_r$ and $s_r$ are the mean and standard deviation of the
  log-normal radius distribution; $R_\mathrm{trunc}$ are the hard truncation bounds on
  the sampled radii.
}
\label{tab:synth_cases}
\scriptsize
\begin{tabular}{llrrrr}
\toprule
Parameter & Unit & Case 1 & Case 2 & Case 3 & Case 4 \\
\midrule
$w_q$                   &             & [0.70,\,0.30] & [0.80,\,0.20] & [0.45,\,0.55] & [0.55,\,0.45] \\
$m_q$                   & \AA$^{-1}$  & [0.100,\,0.110] & [0.100,\,0.140] & [0.100,\,0.140] & [0.070,\,0.075] \\
$s_q$                   & \AA$^{-1}$  & [0.015,\,0.100] & [0.020,\,0.020] & [0.015,\,0.020] & [0.014,\,0.100] \\
Shell thickness         & \AA         & 60   & 65   & 65   & 50   \\
Relative shell contrast &             & 0.96 & 1.05 & 0.85 & 0.98 \\
GRF shift               &             & $-0.1$ & $-0.7$ & $-0.1$ & $-0.3$ \\
$m_r$               & \AA         & 245  & 190  & 190  & 260  \\
$s_r$              & \AA         & 30   & 65   & 25   & 50   \\
$R_\mathrm{trunc}$      & \AA         & [100,\,500] & [100,\,500] & [100,\,500] & [100,\,500] \\
\bottomrule
\end{tabular}
\end{table}

\section{Analytical baseline models}
\label{sec:s8_baselines}

To assess when the heterogeneous core--shell model is needed and what happens to the inferred size distribution when simpler models are used, we fit two analytical baselines to the experimental SAXS curve: a polydisperse homogeneous sphere and a polydisperse core--shell sphere.
All three models (sphere, core--shell, and heterogeneous core--shell) are fitted at three $q$ cutoffs ($q_{\max} \in \{0.02,\, 0.05,\, 0.10\}$~\AA$^{-1}$).
The upper limit of $0.10$~\AA$^{-1}$ is chosen because interior structure contributes substantially above this point and neither analytical model has physics for that regime.

\subsection{Form factors}

\paragraph{Homogeneous sphere.}
The monodisperse form factor is
\begin{equation}
I_{\mathrm{sphere}}(q,r)
= r^6 \bigl[f(qr)\bigr]^2,
\qquad
f(x) = \frac{3(\sin x - x\cos x)}{x^3},\quad f(0)=1.
\end{equation}
Free parameters: mean $\bar{R}$ and standard deviation $\sigma_R$ of a log-normal radius distribution.

\paragraph{Core--shell sphere.}
The monodisperse amplitude is
\begin{equation}
F(q, R, d, c)
= R^3\,f(qR) + r_c^3\,(c-1)\,f(q r_c),
\qquad
r_c = \max(R - d,\, 1~\text{\AA}),
\label{eq:coreshell_amp}
\end{equation}
with $I_{\mathrm{cs}}(q) = F(q)^2$.
Here $R$ is the outer radius, $d$ is the shell thickness, and $c = \Delta\rho_{\rm core}/\Delta\rho_{\rm shell}$ is the contrast ratio, with $\Delta\rho_{\rm shell} \equiv 1$ since the overall scale cancels under Porod-invariant normalization.
Free parameters: $\bar{R}$, $\sigma_R$ (polydisperse, integrated by quadrature), and $d$, $c$ (scalar per restart, not polydisperse).

Both form factors are implemented analytically in PyTorch, with a Taylor expansion near $qr = 0$ to avoid numerical instability.
Polydispersity is incorporated using the same differentiable quadrature layer as in the heterogeneous core--shell model: for each restart, Gauss--Legendre nodes are drawn from the truncated log-normal radius distribution, the monodisperse form factor is evaluated at each node, and the ensemble-averaged intensity is obtained via weighted summation.

\subsection{Fitting procedure}

\paragraph{Shared optimization framework.}
All three models are fitted using PyTorch automatic differentiation, following the same multi-start gradient-based procedure.
All parameters are reparameterized to be bounded and unconstrained in the optimizer's coordinate system, so no projection or penalty terms are required.
Each fitting run uses 4096 restarts, initialized by quasi-random Sobol sampling over the parameter space.
Optimization proceeds with AdamW (learning rate $10^{-2}$, weight decay 0), a cosine annealing schedule over 1000 steps, and gradient clipping at maximum norm 5.0.
Restarts are processed in batches of 256 for GPU efficiency.

\paragraph{Learned parameters and reparameterizations.}
For the sphere model, two parameters are learned per restart: the mean $\bar{R}$ and coefficient of variation $\mathrm{CV} = \sigma_R / \bar{R}$ of the radius distribution, each mapped from an unconstrained raw value via a sigmoid reparameterization to the ranges $\bar{R} \in [150, 350]$~\AA{} and $\mathrm{CV} \in [0.05, 0.50]$.

The core--shell model learns four parameters per restart: the same two size parameters as for the sphere, plus the shell thickness $d$ and the contrast ratio $c$.
These are mapped via $\tanh$ reparameterizations to $d \in [20, 100]$~\AA{} and $c \in [0.5, 2.0]$.
The shell parameters $d$ and $c$ are scalar per restart; they do not vary across the quadrature nodes used for the radial integration.

\paragraph{Loss function.}
Both analytical models are optimized with a composite loss evaluated on $q \leq q_{\max}$:
\begin{equation}
\mathcal{L} = 0.1\,\mathcal{L}_{\mathrm{mse,log}} + \mathcal{L}_{\mathrm{mse,lin}},
\end{equation}
matching the loss used for the heterogeneous core--shell model in the main text (Eq.~11).
All three models use the same composite loss, ensuring that differences in inferred parameters reflect the forward model rather than the fitting objective.

\paragraph{Normalization.}
Before computing the loss, both the predicted and experimental intensities are normalized by the $q^2$-weighted integral evaluated over the fitted window only:
\begin{equation}
\hat{I}(q) = \frac{I(q)}{\int_{q \leq q_{\max}} q^2\,I(q)\,\mathrm{d}q}.
\end{equation}
This normalization is performed identically for all three models, ensuring that differences in fit quality reflect model expressiveness rather than intensity calibration.

\paragraph{Fitted $q$ ranges.}
The choice of $q_{\max}$ controls which structural features are accessible:
\begin{itemize}
  \item $q_{\max} = 0.02$~\AA$^{-1}$: only particle size ($\bar{R}$, $\sigma_R$) is constrained. The shell feature at $q \sim \pi/d \approx 0.05$~\AA$^{-1}$ (for $d \approx 60$~\AA) lies outside this range, so $d$ and $c$ are unconstrained.
  \item $q_{\max} = 0.05$~\AA$^{-1}$: the lower edge of the shell-sensitive regime. The shell thickness $d$ begins to be constrained, but weakly.
  \item $q_{\max} = 0.10$~\AA$^{-1}$: $d$ is well within the fitted range and should be reliably recovered by the core--shell model. At this $q$ the interior structure also contributes, and the core--shell model will show systematic residuals in the $0.05$--$0.10$~\AA$^{-1}$ region where it has no interior physics.
\end{itemize}

Full curves over the entire $q$ range are saved for every restart to enable post-hoc comparison regardless of the fitting window.

\subsection{Results}

\begin{figure}[ht]
    \centering
    \includegraphics[width=\linewidth]{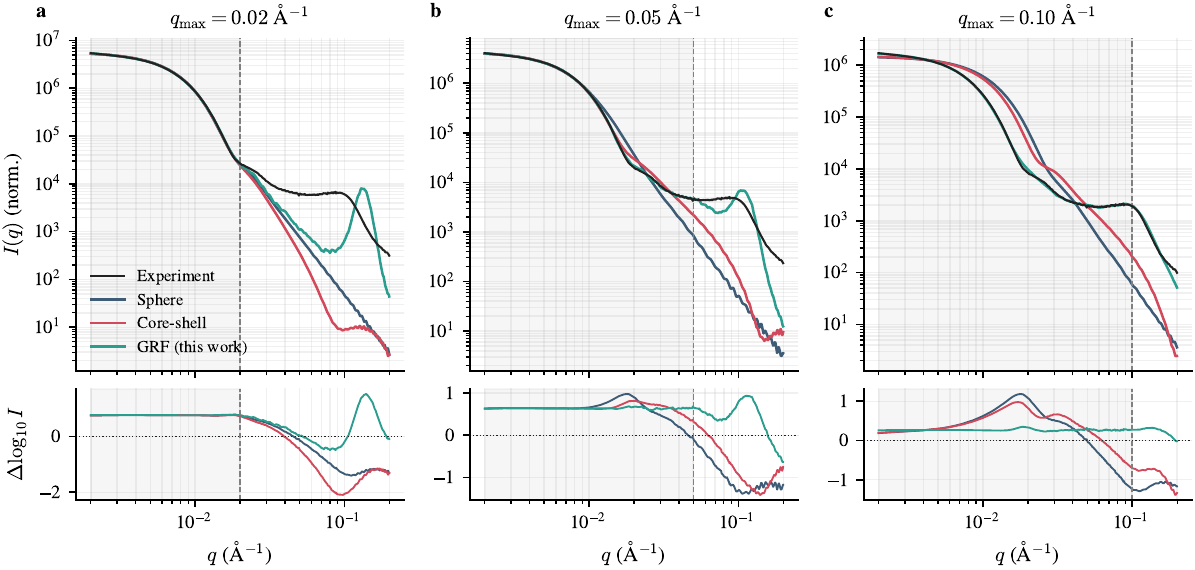}
    \caption{
        \textbf{Fitted curves for sphere, core--shell, and heterogeneous core--shell models at three $q$ cutoffs.}
        Each column corresponds to one fitting range
        ($q_{\max} = 0.02$, $0.05$, $0.10$~\AA$^{-1}$; shaded band).
        Top row: experimental data and best-fit curves shown over the complete
        experimental $q$ range.
        Bottom row: residuals in $\log_{10}$-intensity space.
    }
    \label{fig:model_curves}
\end{figure}

\paragraph{Fitted curves.}
Figure~\ref{fig:model_curves} shows the best-fit curves for each model overlaid on the experimental data, plotted over the full $q$ range, even when optimization used only a subset.
At $q_{\max} = 0.02$~\AA$^{-1}$, all three models reproduce the low-$q$ intensity profile with similar fidelity.
The data in this range constrain particle size but carry little information about shell geometry or interior structure.
At $q_{\max} = 0.05$~\AA$^{-1}$, the homogeneous sphere begins to deviate where the shell modulation enters ($q \sim \pi/d \approx 0.05$~\AA$^{-1}$ for $d \approx 60$~\AA), while the core--shell and heterogeneous core--shell models track the experimental curve more closely.
At $q_{\max} = 0.10$~\AA$^{-1}$, neither the sphere nor the core--shell model can account for the interior-heterogeneity features at intermediate $q$, and both show systematic residuals in the $0.05$--$0.10$~\AA$^{-1}$ region.
The heterogeneous core--shell model maintains low residuals across the full range at all three cutoffs.

\begin{figure}[ht]
    \centering
    \includegraphics[width=\linewidth]{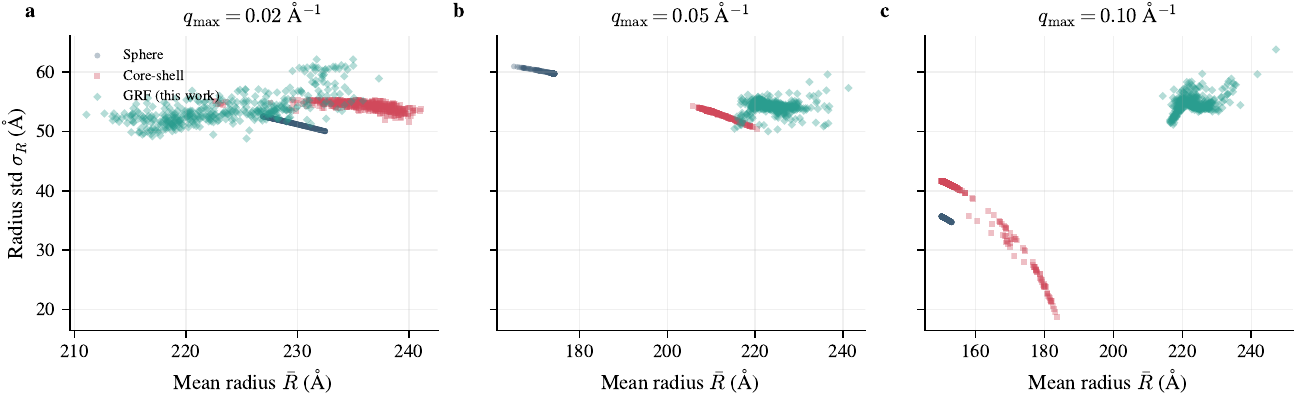}
    \caption{
        \textbf{Inferred size-distribution parameters across models and $q$ cutoffs.}
        Each panel shows a scatter of ($\bar{R}$, $\sigma_R$) for the top 10\%
        of optimization restarts.
        Vertical bands indicate DLS (yellow) and cryo-TEM (purple) mean $\pm$ std.
        The heterogeneous core--shell model parameters remain stable across cutoffs, while the
        analytical models shift to smaller radii as they are forced to fit
        features outside their physics.
    }
    \label{fig:params_scatter}
\end{figure}

\paragraph{Inferred parameters.}
Figure~\ref{fig:params_scatter} compares the size-distribution parameters recovered by each model.
At $q_{\max} = 0.02$~\AA$^{-1}$, all models infer comparable mean radii ($\bar{R} \approx 220$--$240$~\AA) and similar polydispersity.
This is expected: the low-$q$ portion of the curve is dominated by particle geometry, and any model with the right number of size-distribution degrees of freedom can capture it.
At this cutoff, the core--shell parameters $d$ and $c$ are not constrained by the data, so their values are not reported.

As $q_{\max}$ increases to 0.05 and 0.10~\AA$^{-1}$, the analytical models are forced to fit features that their physics cannot produce.
The sphere model compensates by shifting $\bar{R}$ downward and adjusting $\sigma_R$, resulting in size estimates that are increasingly inconsistent with the independent DLS and cryo-TEM measurements.
The core--shell model shows a similar, though less severe, drift.
The heterogeneous core--shell model maintains stable size-distribution parameters across all three cutoffs, because its spectral parameters absorb the interior contribution without distorting the radius distribution.

\paragraph{Size-distribution stability and apparent precision.}
Figure~\ref{fig:params_scatter} also shows the inferred mean radius and standard deviation for the top 10\% of fits from each model, plotted as a scatter across the three $q$ cutoffs.
At $q_{\max} = 0.02$~\AA$^{-1}$, the three model ensembles overlap substantially, confirming that size information alone does not differentiate the models.
At higher cutoffs, the sphere and core--shell ensembles migrate to smaller mean radii and shifted size distributions, reflecting the compensatory adjustments described above.
The heterogeneous core--shell ensemble remains in a consistent region across all cutoffs, and its inferred radii are closer to the DLS and cryo-TEM values than the analytical models at $q_{\max} \geq 0.05$~\AA$^{-1}$.

A related observation concerns the spread of accepted solutions at low $q$, where all three models fit the curve comparably.
Even at $q_{\max} = 0.02$~\AA$^{-1}$, the range of mean radii across near-optimal solutions differs between models.
The sphere model accepts a narrow band (approximately $\bar{R} \in [227, 233]$~\AA), the core--shell model a somewhat wider range ($\bar{R} \in [225, 242]$~\AA), and the heterogeneous core--shell model the widest ($\bar{R} \in [210, 245]$~\AA).
This is because interior heterogeneity slightly modulates the scattering even at low $q$, and the models without interior degrees of freedom cannot accommodate this variation.
The result is that simpler models appear to constrain the size distribution more precisely, but that apparent precision is an artifact of missing physics: the narrower range excludes solutions that a more complete model considers plausible.

\begin{figure}[ht]
    \centering
    \includegraphics[width=0.65\linewidth]{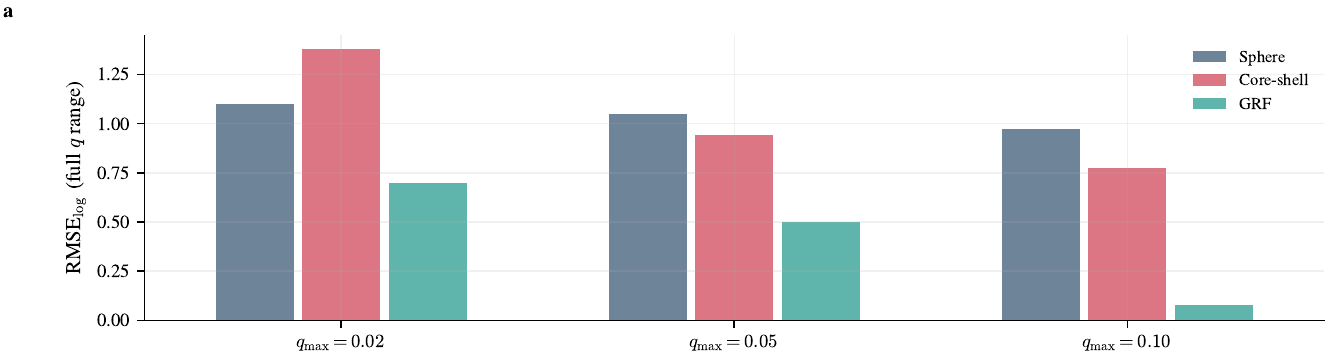}
    \caption{
        \textbf{RMSE$_{\log}$ on the full $q$ range for each model and fitting cutoff.}
        The heterogeneous core--shell model maintains low prediction error across all cutoffs,
        while the analytical models degrade as interior-structure features enter the fitted range.
    }
    \label{fig:rmse_full}
\end{figure}

\paragraph{RMSE on the full $q$ range.}
Figure~\ref{fig:rmse_full} reports $\mathrm{RMSE}_{\log}$ evaluated on the complete $q$ range for each model and cutoff, regardless of where the fit was performed.
This is a post-hoc extrapolation metric: a model that fits its target region well but lacks physics outside it will show high full-range RMSE.
All three models achieve low RMSE when fitted only to $q \leq 0.02$~\AA$^{-1}$, but the sphere and core--shell errors grow substantially at higher cutoffs.
The heterogeneous core--shell model maintains low full-range RMSE at all three cutoffs, consistent with its ability to represent both particle-scale and interior-scale contributions to the scattering curve.

\paragraph{Summary.}
The comparison confirms that the heterogeneous core--shell model earns its additional complexity.
At low $q$, where only particle size is probed, all three models perform comparably in terms of curve fit, and a simpler model may be preferred for transparency.
However, even at low $q$ the simpler models give a misleadingly narrow range of plausible size-distribution parameters.
Once the fitting range includes shell or interior features, the analytical models either fail to fit (high RMSE) or absorb the misfit into biased size-distribution parameters.
The heterogeneous core--shell model avoids this because it has explicit degrees of freedom for the interior electron-density structure, yielding both better curve agreement and more physically consistent size estimates across the full measured $q$ range.

\subsection{Structured core--shell baseline with analytical correlation peak}
\label{sec:structured_coreshell}

To test whether the interior scattering features captured by the GRF model can also be reproduced by a conventional analytical form factor, we fit a third baseline: a polydisperse core--shell model augmented with an explicit Gaussian correlation peak inside the core. Unlike the automated multi-start procedure used for the other baselines and for the GRF model, this fit required manual parameter tuning and took several hours to converge, illustrating the practical difficulty of extending analytical models to heterogeneous interiors.

\paragraph{Model.}
The total intensity is the size-distribution average of the single-particle intensity $I_{sp}(q, r)$:
\begin{equation}
I(q) = \frac{1}{\mathcal{N}} \int_{0}^{\infty} D(r)\, I_{sp}(q, r) \, dr
\label{eq:Analytical_fitting}
\end{equation}
where $\mathcal{N}$ is a normalization constant and $r$ is the core radius.
Each particle's intensity is the sum of a coherent core--shell envelope and an incoherent interior contribution:
\begin{equation}
I_{sp}(q, r) = \left| A_{\mathrm{bulk}}(q, r) \right|^2 + I_{\mathrm{int}}(q, r).
\end{equation}
The bulk amplitude is defined by the contrast differences between the core ($\rho_c$), shell ($\rho_s$), and solvent ($\rho_v$):
\begin{equation}
A_{\mathrm{bulk}}(q, r) = (\rho_s - \rho_v)\, V(r+t)\, \Phi(q, r+t) + (\rho_c - \rho_s)\, V(r)\, \Phi(q, r)
\end{equation}
where $t$ is the shell thickness, $V(R) = \frac{4}{3}\pi R^3$, and $\Phi(q, R) = 3[\sin(qR) - qR\cos(qR)]/(qR)^3$ is the spherical form factor.
The interior contribution models phase separation within the core as a Gaussian structure factor windowed by the core form factor:
\begin{equation}
I_{\mathrm{int}}(q, r) = P_{\mathrm{weight}} \cdot \exp\!\left( -\tfrac{1}{2} \left[ \frac{q - q_0}{\gamma} \right]^2 \right) \cdot \left[ V(r)\, \Phi(q, r) \right]^2
\end{equation}
where $q_0 = 2\pi / d_{\mathrm{spacing}}$ is the peak position, $\gamma$ is the peak width, and $P_{\mathrm{weight}}$ controls the relative strength of the interior scattering.
The size distribution $D(r)$ is log-normal with mean radius $\bar{r}$ and polydispersity $PD = \sigma/\bar{r}$:
\begin{equation}
D(r) = \frac{1}{r \,\sigma_{\ln} \sqrt{2\pi}} \exp\!\left( -\frac{(\ln r - \mu)^2}{2\sigma_{\ln}^2} \right)
\label{eq:analytic_polydispersity}
\end{equation}
where $\sigma_{\ln} = \sqrt{\ln(1+PD^2)}$ and $\mu = \ln(\bar{r}) - \tfrac{1}{2}\sigma_{\ln}^2$.

\begin{figure}[t!]
    \centering
    \includegraphics[width=0.74\linewidth]{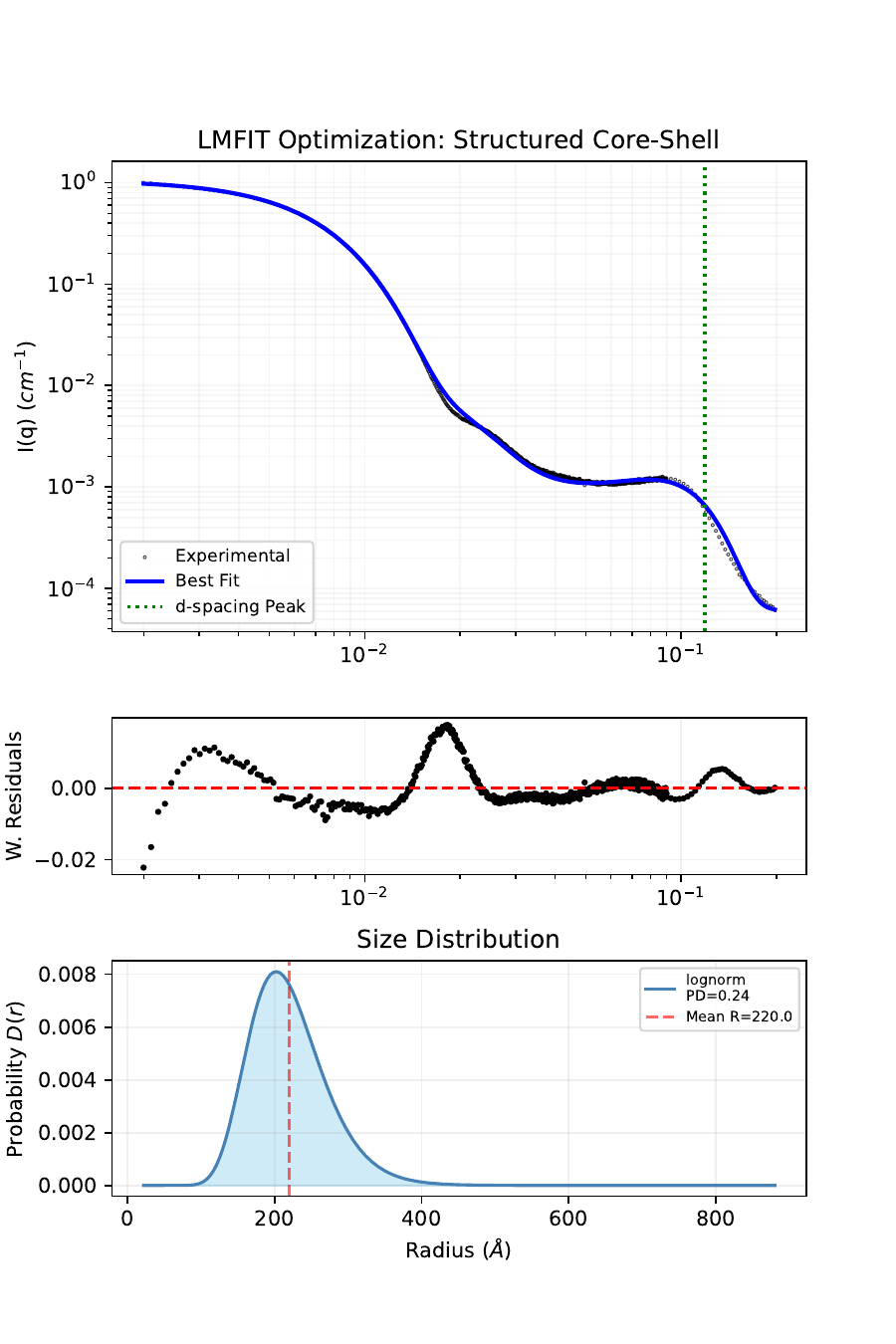}
    \caption{\textbf{Structured core--shell baseline fit.} (a)~Best fit of Eq.~\ref{eq:Analytical_fitting} to the experimental MC3 LNP data (cf.\ Fig.~4(d) in the main text); the vertical line marks the periodicity of the interior correlation peak. (b)~Residuals between the best fit and the experimental data. (c)~Fitted log-normal polydispersity distribution (Eq.~\ref{eq:analytic_polydispersity}).}
    \label{fig:Analytical_fitting}
\end{figure}

\paragraph{Result.}
Figure~\ref{fig:Analytical_fitting} shows that the structured core--shell model can reproduce the experimental SAXS curve, including the interior scattering feature around $q \approx 0.06$~\AA$^{-1}$. However, this required manual selection of the peak parameters ($q_0$, $\gamma$, $P_{\mathrm{weight}}$) and iterative adjustment over several hours. By contrast, the GRF parameterization captures the same features automatically within the multi-start fitting framework, and the resulting ensemble of near-optimal solutions provides information about parameter trade-offs and identifiability that a single manually tuned fit cannot.

\section{TEM and DLS characterization}
\label{sec:s9_tem_dls}

Sample preparation, instrumentation, and measurement conditions for both cryo-TEM and DLS are described in the Methods section of the main text.
This section provides the supporting figures.

\subsection{Transmission electron microscopy}
\begin{figure}[t!]
    \centering
    \includegraphics[width=8cm]{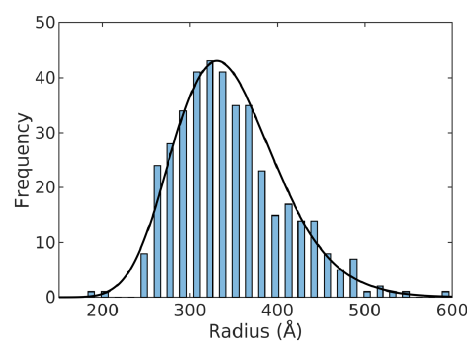}
    \caption{
    \textbf{TEM particle size histogram.}
    Equivalent radii from manual annotation of 400 particles, with a log-normal fit superimposed.
    Mean radius $346 \pm 62$~\AA.
    }
    \label{fig:tem_histogram}
\end{figure}

Figure~\ref{fig:tem_histogram} shows the distribution of equivalent radii from manual annotation of 400 particles, with a log-normal fit superimposed.

\subsection{Dynamic light scattering}

\begin{figure}[t!]
    \centering
    \includegraphics[width=16cm]{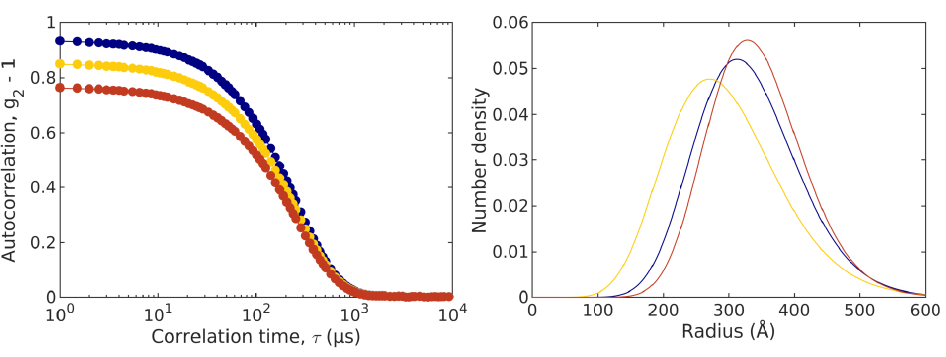}
    \caption{
    \textbf{DLS analysis.}
    Fits to the intensity autocorrelation function $g_2(\tau)$ (left); the corresponding Schultz number-weighted size distribution (right).
    Mean hydrodynamic radius $323 \pm 79$~\AA{} (triplicate average).
    }
    \label{fig:dls}
\end{figure}

\newpage
\section{PyTorch implementation}

\begin{lstlisting}[language=Python, caption={DCT surrogate model implementation.}]
import torch
import torch.nn as nn
import numpy as np
from scipy.fft import dct

def make_dct2_ortho_uniform(N_in, N_out):
    # Build orthonormal DCT-II basis matrix of shape (N_out, N_in)
    D = dct(np.eye(N_in), type=2, norm='ortho', axis=0)
    return D[:N_out, :]

class DCTModel(nn.Module):
    def __init__(self, Phi_dct, n_layers=9, hidden=1024):
        super().__init__()
        self.register_buffer('Phi', Phi_dct)  # (N_q-1, N_q)
        layers = []
        in_dim = 10  # theta dimension
        for _ in range(n_layers - 1):
            layers += [nn.Linear(in_dim, hidden), nn.SiLU()]
            in_dim = hidden
        layers += [nn.Linear(hidden, Phi_dct.shape[0] + 2)]  # mu, log_sigma, coeffs
        self.net = nn.Sequential(*layers)

    def forward(self, theta):
        out      = self.net(theta)
        mu       = out[:, 0:1]
        log_std  = out[:, 1:2]
        z_ac     = out[:, 2:]
        core_std = z_ac @ self.Phi
        return mu + log_std.exp() * core_std

Phi_dct = torch.from_numpy(make_dct2_ortho_uniform(300, 300)).float()
model   = DCTModel(Phi_dct, 9, hidden=1024).to("cuda")
\end{lstlisting}

\end{document}

%% file: math_commands.tex
\usepackage{amsmath,amsfonts,bm}

\def\eqref#1{equation~(\ref{#1})}

\def\1{\bm{1}}

\DeclareMathAlphabet{\mathsfit}{\encodingdefault}{\sfdefault}{m}{sl}
\SetMathAlphabet{\mathsfit}{bold}{\encodingdefault}{\sfdefault}{bx}{n}

